\documentclass[preprint]{aastex}
\newcommand{\HI}{H~{\sc i}} 

\newcommand{\kms}{${\rm km~s^{-1}}$}

\slugcomment{To Appear in ApJS}
\shortauthors{McCLURE-GRIFFITHS ET AL.} 
\shorttitle{GASS I. Description, Goals and Data}

\begin{document} 

\title{GASS: The Parkes Galactic All-Sky Survey. I. Survey Description, Goals, and
 Initial Data Release}

\author{N.\ M.\ McClure-Griffiths,\altaffilmark{1} D.\ J.\
  Pisano,\altaffilmark{2,3} M.\ R.\ Calabretta,\altaffilmark{1} 
  H.\ Alyson Ford,\altaffilmark{1,4} Felix J.\
  Lockman,\altaffilmark{2} L.\ Staveley-Smith,\altaffilmark{5} P.\ M.\ W.\
  Kalberla,\altaffilmark{6} J.\ Bailin,\altaffilmark{7} L.\
  Dedes,\altaffilmark{6} S.\ Janowiecki,\altaffilmark{2,8} B.\ K.\
  Gibson,\altaffilmark{9} T.\ Murphy,\altaffilmark{10,11} H.\
  Nakanishi,\altaffilmark{12} K.\ Newton-McGee\altaffilmark{1,10}}

\altaffiltext{1}{Australia Telescope National Facility, CSIRO, Marsfield
  NSW 2122, Australia; naomi.mcclure-griffiths@csiro.au, mark.calabretta@csiro.au}
\altaffiltext{2}{National Radio Astronomy Observatory, Green
  Bank, WV 24944; dpisano@nrao.edu, jlockman@nrao.edu}

\altaffiltext{3}{present address: Department of Physics, West Virginia
  University, Morgantown, WV 26506}

\altaffiltext{4}{Centre for Astrophysics and Supercomputing, Swinburne
  University of Technology, Hawthorn VIC 3122, Australia;
  alyson@astro.swin.edu.au}

\altaffiltext{5}{School of Physics, University of Western Australia,
  Crawley WA 6009, Australia; lister.staveley-smith@uwa.edu}

\altaffiltext{6}{Argelander-Institut f\"ur Astronomie, Universit\"at
  Bonn, 53121 Bonn, Germany; pkalberla@astro.uni-bonn.de,
  ldedes@astro.uni-bonn.de}

\altaffiltext{7}{Department of Physics and Astronomy, McMaster
  University, Hamilton, ON L8S 4M1, Canada; bailinj@mcmaster.ca}

\altaffiltext{8}{Department of Astronomy, Case Western Reserve
  University, Cleveland, OH 44106; present address: Department of
  Astronomy, Indiana University Bloomington, IN 47405;
  sjanowie@indiana.edu}

\altaffiltext{9}{Centre for Astrophysics, University of Central
  Lancashire, Preston, PR1 2HE, UK; bkgibson@uclan.ac.uk}

\altaffiltext{10}{School of Physics, The University of Sydney, NSW 2006,
  Australia; tara@physics.usyd.edu.au}

\altaffiltext{11}{School of Information Technologies, The University of
  Sydney, NSW 2006, Australia}

\altaffiltext{12}{Faculty of Science, Kagoshima University, Kagoshima
  890-0068, Japan; hnakanis@sci.kagoshima-u.ac.jp}

%--------------------------------------------
\begin{abstract}
%-------------------------------------------
  The Parkes Galactic All-Sky Survey (GASS) is a survey of Galactic
  atomic hydrogen (\HI) emission in the Southern sky covering
  declinations $\delta \leq1\arcdeg$ using the Parkes Radio Telescope.
  The survey covers $2\pi$ steradians with an effective angular
  resolution of $\sim 16\arcmin$, at a velocity resolution of $1.0$
  \kms, and with an rms brightness temperature noise of $57$ mK.  GASS
  is the most sensitive, highest angular resolution survey of Galactic
  \HI\ emission ever made in the Southern sky. In this paper we
  outline the survey goals, describe the observations and data
  analysis, and present the first-stage data release.  The data
  product is a single cube at full resolution, not corrected for stray
  radiation.  Spectra from the survey and other data products are
  publicly available online.
\end{abstract}

%----------------------------------------------------------
\section{Introduction}
\label{sec:intro}
%----------------------------------------------------------
Atomic hydrogen (\HI) is a ubiquitous component of disk galaxies.
Most readily traced by the $\lambda = 21$ cm spectral line, \HI\ from
the Milky Way is observed in all directions of the sky.  It is
possible to trace Galactic \HI\ emission to the far side of the Milky
Way, probing Galactic and interstellar processes in regions of the
Galaxy that are inaccessible at many other wavelengths.  As the
dominant component of the interstellar medium (ISM) by number, \HI\
allows us to trace a wide variety of Galactic processes including the
impact of massive stars on the ISM
\citep[e.g.,][]{heiles84,mcgriff02a}; the interaction of the Galactic
disk and halo \citep[e.g.,][]{mcgriff06a,lockman08}; the ISM
life-cycle; and the formation of cold clouds
\citep[e.g.,][]{gibson00,kavars05}. Since the discovery of the
$\lambda=21$ cm spectral line in 1951 by \citet{ewen51}, it has been
used repeatedly, and with continuous refinements, to explore the
rotation curve and map the global structure of the Galaxy
\citep[e.g.,][]{kerr62,henderson82,levine06,mcgriff07}.

Not long after the discovery of the \HI\ spectral line, \HI\ emission
was found at velocities in excess of $|V_{LSR}|\gtrsim 100$ \kms\ that
could not be explained by Galactic rotation \citep{muller63,smith63}.
These high velocity clouds (HVCs) are now known to cover a significant
fraction of the sky \citep{wakker97}.  HVCs are observed at both
positive and negative LSR velocities with magnitudes up to $|V_{LSR}|
\lesssim 500$ \kms.  There have been extensive searches for high
velocity gas (see Wakker \& van Woerden 1997 for a review, or recent
surveys by Putman et al.\ 2002 and Lockman et al.\ 2002b).
\nocite{putman02,lockman02b,wakker97} It now seems certain that HVCs
represent a variety of phenomena.  Some HVCs may be related to a
Galactic fountain \citep[e.g.,][]{bregman80}; some are tidal debris,
such as those connected to the Magellanic Stream
\citep[e.g.,][]{putman03} or other satellites
\citep[e.g.,][]{lockman03}; some may be infalling intergalactic gas
\citep[e.g., Complex C;][]{wakker99,tripp03,lockman08}l; and some may
be associated with gas condensing from a massive, hot halo
\citep{maller04,sommer-larsen06}.  Studies of the structure and
distribution of high velocity gas provide critical information on the
evolution of the Milky Way system.

All-sky surveys of Galactic \HI\ with broad bandwidths and high
spectral and angular resolution allow astronomers to simultaneously
explore Milky Way structure, the ISM and HVCs.  The most recent such
survey is the Leiden-Argentine-Bonn survey \citep[LAB;][]{kalberla05},
which has produced a sensitive ($\sigma\ T_B = 70-90$ mK) database of
the entire sky with a beam size of $\sim 30\arcmin$ sampled on a
$30\arcmin$ grid, giving an effective resolution of $\sim 1\arcdeg$.
However, we know from high resolution studies of targeted areas of the
Galactic halo \HI\ emission that there is a wealth of interesting
structure on scales smaller than 30\arcmin\
\citep{lockman02a,stanimirovic06,peek07}.  For example, Green Bank
Telescope (GBT) results from \citet{lockman02a} show that the halo of
the Galaxy contains a significant population of small, cold clouds,
possibly products of the Galactic fountain, that are not detectable
with the angular resolution of the LAB survey.  Similarly, high
angular and spectral resolution studies have resolved physical and
spectral structure in many HVCs, giving clues to their nature
\citep{bruens05,westmeier05}.  The  HIPASS survey (Barnes et
al.\ 2001, Putman et al.\ 2002)\nocite{barnes01,putman02} covered the
sky at $\sim 15\arcmin$ resolution, but was not designed for Galactic
\HI, and therefore did not accurately measure low velocity gas, nor
have the velocity resolution needed to resolve interstellar \HI\
lines.  While high angular resolution Galactic Plane surveys
\citep{taylor03,mcgriff05,stil06} have allowed us to carefully explore
Galactic \HI\ emission in the disk, they are not very sensitive
($\sigma\ T_B \sim 1$ K) and are restricted to a few degrees around
the Galactic Plane, leaving most of the volume of the Galaxy
unexplored on scales less than a degree.  To fully understand the
nature and origin of \HI\ structure in the halo and of high velocity
clouds there is a need for a sensitive, unbiased high resolution
survey of the entire sky.

We have made a new sensitive, fully sampled, high resolution sky survey
of Galactic \HI\ emission south of declination $\delta = 1\arcdeg$,
with the Parkes 64 m radio telescope:  the Galactic All-Sky Survey (GASS).  
It was designed with the primary goals of studying
the interaction of the Milky Way disk and halo and the nature of HVCs
and intermediate velocity clouds \citep[IVCs; e.g.,][]{kuntz96}.
Secondary science goals include Milky Way structure and the extended
structure of the Magellanic system.  Here we describe the GASS
project, focusing on the survey goals and techniques. In \S
\ref{sec:goals} we describe the main design goals of the survey
and the details about the observations and data reduction techniques
are given in \S \ref{subsec:obs} and \S \ref{subsec:data},
respectively.  In \S \ref{sec:products} we present the data
products and discuss the limitations of this first stage of our 
two-stage data release.  Finally, in \S \ref{sec:disc} we briefly discuss 
some applications of the first release data and describe some initial GASS
results.

%----------------------------------------------------------
\section{Observations and Data Reduction}
\label{sec:details}
%----------------------------------------------------------

\subsection{Survey Design}
\label{sec:goals}

The parameters of the survey were chosen to match specific criteria
relating to the spatial and velocity extents and velocity widths of
known \HI\ emission.  Accurate measurement of Galactic \HI\ at all
velocities requires that the data be taken by frequency-switching, as
\HI\ emission, especially at $V_{LSR} \approx 0$ \kms, is always very
extended.  The survey had to cover at least $-400 < V_{LSR} < +400$
\kms, which is the range of high-velocity \HI\ in the Southern
hemisphere \citep{putman02}.  Throughout this work all velocities are
given in the kinematic Local Standard of Rest (LSRK), also commonly
referred to simply as LSR, and defined from an average of the
velocities of stars in the Solar neighbourhood
\citep{delhaye65,gordon76}. The velocity resolution had to be
sufficient to measure the narrow lines (a few \kms) found in HVC cores
and disk \HI\ \citep{haud07,kalberla06,lockman02b, bruens00}.  The
sensitivity goal was set to match or exceed the LAB survey, but with
higher angular resolution.  The survey completely covers the sky with
Nyquist sampling so that the data can be used to provide
short-spacings for interferometric imaging.

GASS was designed to meet these criteria by fully sampling Galactic
\HI\ in the velocity range $-400 \leq v \leq 500$ \kms\ over the sky
south of $\delta = 1\arcdeg$ with angular resolution of $16\arcmin$
and a velocity resolution of $1$ \kms.  The survey used in-band
frequency-switching, described below, to maximize observing efficiency
while preserving sensitivity to extended emission.  Integration times
were chosen to achieve a brightness temperature noise ($1\sigma$) $\le
70$ mK per channel thus matching the LAB sensitivity for extended
sources while having four times the sensitivity of the LAB survey for
unresolved sources.

\subsection{Observations}
\label{subsec:obs}

GASS observations were made with the 21 cm multibeam receiver system
on the Parkes 64 m radio telescope.  The 21 cm multibeam has 13 dual
linear polarization receivers in the single cooled dewar.  The 13 feed
horns are arranged in a hexagonal pattern at the focal plane,
consisting of a central feed, an inner ring of six feeds and an outer
ring with an additional six feeds \citep[see][]{staveley-smith96}. At
$\lambda =21$ cm, the average measured beamwidths for the multibeam
are: $14.0\arcmin$ for the central beam, $14.1\arcmin$ for beams two
to seven with an ellipticity of 0.03, and $14.5\arcmin$ for beams
eight to 13 with an ellipticity of 0.06 \citep{staveley-smith96}.  The
mean beamwidth is therefore $14.3\arcmin$ with a separation between
adjacent beam centers of $29.1\arcmin$.  If the receiver is oriented at an 
angle of $19.1\arcdeg$ with respect to the scan direction
it produces equally spaced tracks with the inner seven beams.

Observations were conducted between 28 January 2005 and 1 November
2006 in eight observing sessions of typically two weeks in duration.
Observing sessions were organised to cover a contiguous region of the
sky, usually several hours in right ascension and all declinations.
All observations were conducted at night and at elevations greater
than $30\arcdeg$, which is the Parkes elevation limit.

GASS consists of two complete surveys of the sky: one scanned in
declination, one scanned in right ascension.  The receiver angle was
adjusted every 5~s to ensure that the individual beams of the
multibeam produce approximately equally spaced tracks on the sky
parallel to the scan direction.  GASS was designed to be fully sampled
with only the inner seven beams to simplify future stray radiation
corrections, but data from all 13 beams are used for the results 
presented here.

Spectra were obtained using a special purpose correlator mode,
combining the Multibeam correlator and the Wideband correlator to
achieve 8 MHz of bandwidth divided into 2048 channels for both
polarizations on all 13 beams.  The data were recorded with
``in-band'' frequency-switching, in which spectra are recorded at two
closely spaced frequencies (IFs) centered at 1418.8435 and 1421.9685
MHz with a 5~s duty cycle.  Frequency-switching allows one to
effectively remove the continuum signal and time variations in the
system gain.  In-band frequency switching, where both bands contain
the \HI\ line, has the added benefit of maximising on-source time
because the two spectra can be combined to yield a total about 4.5 MHz or 950
\kms\ of continuous bandwidth.  The channel spacing is $\Delta v = 0.82$
\kms\ and the effective channel width is $1.0$ \kms.  Doppler tracking
was not applied on-line so the LSRK velocity range accessible varies
slightly for each spectrum.  The frequency switch of 3.125 MHz
corresponds to 660 \kms, so every real emission line feature has an
associated, spurious, negative image displaced by $\pm$660 \kms, but
most of these negative images fall outside the velocity coverage of
the survey.  We discuss this further in \S~\ref{sec:artifacts}.

\begin{figure}
\centering
\includegraphics[angle=-90,width=6in]{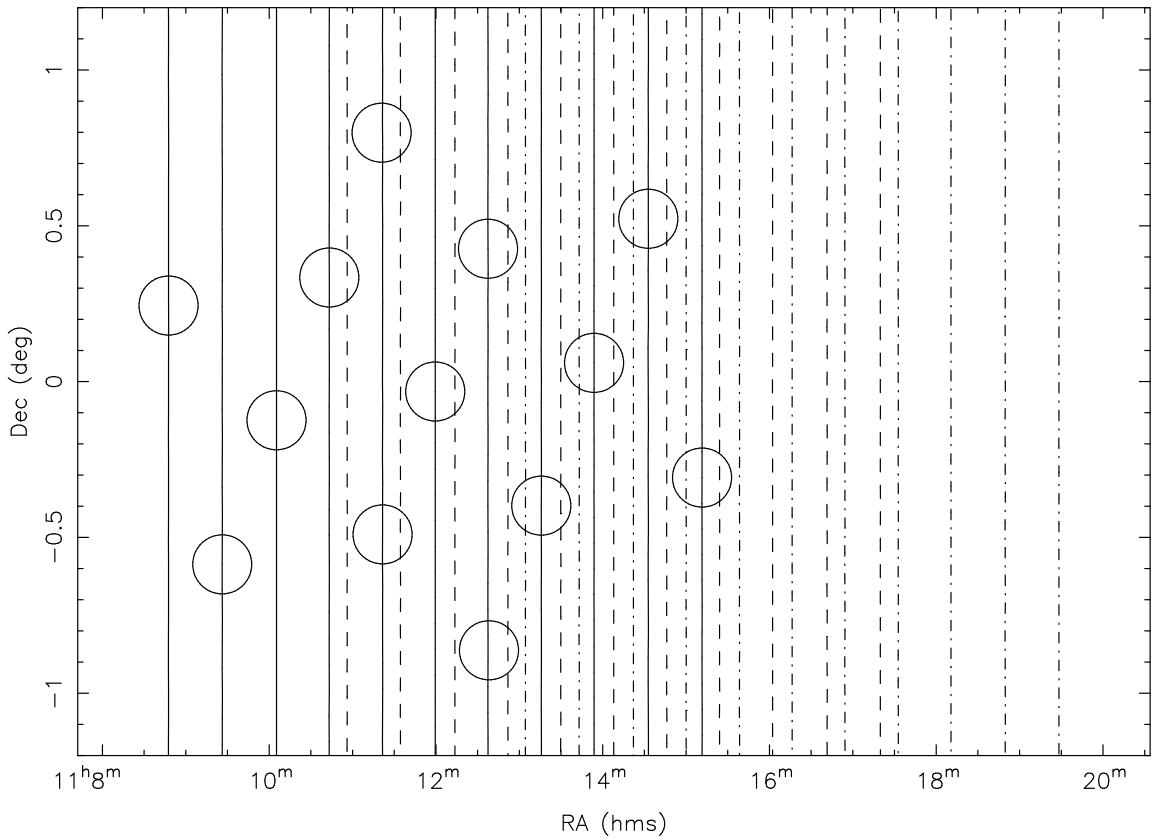}
\caption[]{Scan pattern for three consecutive scans of the multibeam receiver.
  The three scans are shown by the solid line, the dashed line and the
  dash-dotted line.  Overlaid on the first scan are circles showing
  the multibeam hexagonal pattern.  The
  receiver is rotated at $19.1\arcdeg$ with respect to the scan
  direction, which is the optimal angle for a seven beam system.  As a
  result several outer beams track the same part of sky as the inner beams. 
\label{fig:scanpattern}}
\end{figure}

A scan consists of all the data obtained while driving the telescope through 
8 degrees of right ascension or declination at a rate of 
$1~{\rm deg~min^{-1}}$.  It is composed of 26 independent sub-scans from the
13 beams and two polarizations with data samples every 5~s.  With these 
scanning and sampling rates the spacing
between adjacent samples is $5\arcmin$, which compares favourably with
the Nyquist sampling, $\lambda/2D = 5.6\arcmin$ for an observing
wavelength of $\lambda=21$ cm and the dish diameter of $D=64$ m.  The
separation between adjacent beam tracks with a $19.1\arcdeg$ receiver
rotation angle is $9.4\arcmin$.  For Nyquist sampling interleaved scans 
were required.  Consecutive scans were offset by $32\arcmin$ so that
adjacent beam tracks were observed with different receivers.  An example 
of the scanning pattern for
three interleaved scans is shown in Figure~\ref{fig:scanpattern}.
After three scans the spacing between adjacent tracks is $3.1\arcmin$.
On-the-fly observing and subsequent gridding broadens the effective 
telescope beamwidth to $16\arcmin$.  The integration time per spectrum (pixel)
was 30 seconds for the final data product.

%----------------------------------------------------------
\subsection{Data Reduction}
\label{subsec:data}
%----------------------------------------------------------
The majority of the data reduction was carried out with {\em Livedata}
and {\em Gridzilla} software\footnote{Binaries and source code are
available from
\url{http://www.atnf.csiro.au/computing/software/livedata.html}}
specifically designed to process multibeam data.  {\em Livedata}
performs bandpass correction and flux calibration and {\em Gridzilla}
produces gridded images from the calibrated, corrected data.

\subsubsection{Bandpass Correction: GASS Mode of {\em Livedata}}
\label{subsubsec:bandpass}

The {\em Livedata} processing pipeline was developed and used for the
HIPASS survey as described by Barnes et al.\ (2001)\nocite{barnes01}.
In its original form, {\em Livedata} derived a bandpass solution from
the emission-free portions of a scan.  However, neither this nor
similar techniques are suitable for GASS because Galactic \HI\
emission covers the whole sky.  

From its origin, {\em Livedata} established the use of robust
estimation, based mainly on the use of median statistics, and this was
significant in ameliorating the effects of radio frequency interference
(RFI) in the HIPASS survey.  For GASS processing, {\em Livedata} was
adapted to use similar robust techniques for bandpass calibration of
frequency-switched Galactic \HI\ data.  It works with the quotient of
each frequency-switched pair, carefully masking any emission, both
spectrally and in the time domain, before determining the average
bandpass solution for each scan.  The bandpass is calibrated separately
for each beam and polarization.

In this section we describe in detail the procedures used to perform
the bandpass correction for each beam and polarization, including the
formation of quotient spectra, masking emission spectrally and
temporally and fitting the average bandpasses. These steps are also
graphically outlined in Figure~\ref{fig:flowchart}. We first define
some variables.  Let subscripts $i_1$ and $i_2$ denote the first and
second integration in pair number $i$.  The spectrum centred at
1418.8435 MHz is $S_{i_1}(\nu)$, and the next integration,
$S_{i_2}(\nu)$, is centered at 1421.9685 MHz. The first integration,
$S_{i_1}(\nu)$, is divided by $S_{i_2}(\nu)$ to produce a quotient
spectrum $q_{i_1}(\nu)$, and vice versa to produce $q_{i_2}(\nu)$.  In
general we use the variable $q$ to denote quotient spectra, $\hat{q}$
to denote time averages of quotient spectra, and $Q$ to denote
estimations of the baseline, which may be medians or polynomial
fits. Steps in the first and second passes, as described below, are
distinguished by the use of prime such as, $q'$ and $q''$.

Considering the first in each pair of quotients, $q_{i_1}(\nu)$, a
time-averaged value in emission-free regions for the eight minute scan
is determined for each channel.  This is ultimately fit to provide the
final bandpass solution, $Q''(\nu)$.  Because $q_{i_1}(\nu)$ and
$q_{i_2}(\nu)$ are reciprocals, the bandpass solution need only be
computed for one or the other.  Much of the complexity of the
algorithm relates to the identification of line emission and RFI and
the construction of masks that this identification entails.  These
masks are determined iteratively in two passes.  

The first pass begins with the determination of a base level and
characteristic deviation in emission-free regions.  The median
value over $\nu$ of $q_{i_1}(\nu)$ is computed for each $i$ in the scan.
The median of these medians then gives the {\em median of quotients},
$Q_1$, a single number that provides a zeroth-order approximation to the
base-level of $q_{i_1}(\nu)$ over frequency and time.  Normally $Q_1$ will 
be close to unity.  Now for each $i$, we compute the median value over $\nu$ of 
$|q_{i_1}(\nu) - Q_1|$.  The median of these medians is the {\em median 
quotient deviation}, $D_1$.  This is used together with $Q_1$ to identify 
line emission or absorption when selecting data to form a time-average 
quotient value for each channel.
Thus, we consider each channel of $q_{i_1}(\nu)$ in turn for all $i$ in
the scan; i.e.\ as a function of time.  We reject $q_{i_1}(\nu)$ for
statistical purposes if $|q_{i_1}(\nu) - Q_1| > 3 D_1$.  The rejected 
$q_{i_1}(\nu)$ form a {\em time mask} for the spectral channel indicating 
where the scan has passed through a source or encountered transient RFI.

The time mask is then subjected to a broadening algorithm that aims to
exclude the low-level wings of the source profile.  This step
recognizes that masking by means of a discriminant only accounts for
the central part of an emission line or RFI.  The choice of time
masking broadening parameters is a careful balance between defining a
mask that is too small so that the diffuse wings of sources are fit as
bandpass and the source is clipped or defining a mask that is too
large so that many masks merge together and the majority of a scan is
masked.  Many values for the time mask were tested before settling on
parameters such that each patch of consecutive false values (rejected
samples) in the mask is extended by one on each side if it is at least
2 elements wide, and by a further one on each side for every
additional 4 elements.  Thus a single isolated false value in the mask
is left alone on the basis that it is probably a noise spike.
Isolated patches that consist of $1, 2, 3, 4, 5, 6, 7, 8, 9, \ldots$
consecutive false values will broaden to $1, 4, 5, 6, 7, 10, 11, 12,
13, \ldots$ samples.  As the patch size increases the limiting value
of the broadening factor is 150\%.  In this process it is not uncommon
for neighbouring masked regions to blend into a single, larger masked
region.  An example of the first pass time-masked quotient spectrum is
shown in the third panel of Figure~\ref{fig:flowchart}.

After broadening the time mask, or even beforehand, there will usually
be some channels for which too few quotients remain to compute a
meaningful time-average value; typically this occurs at low \HI\
velocities.  If fewer than 90\% of quotients for a given channel are
rejected, a first approximation to the bandpass solution for the
channel, ${\hat q'_1}(\nu)$, is computed as the median value of those
 remaining.  If more than 90\% are rejected, then this channel is
masked for the entire scan forming a {\em channel mask}.  Channel
masks are then subject to the same broadening process as the time
masks, though with more aggressive broadening parameters: the mask is
extended by one channel on each side for the first false value, and by
a further one on each side for every additional 2 false values.  Thus
an isolated patch of $1, 2, 3, 4, 5, 6, \ldots$ false values becomes
$3, 4, 7, 8, 11, 12, \ldots$ with a limiting broadening factor of
200\%.  Once again, these masking values were chosen as a compromise
between channel masks that are so small that real emission in the
line wings is included in the fit and masks that are so large that a
significant fraction of the band is masked and as a consequence the fit is
poorly constrained. {\em Livedata} also allows for the channel mask to
be augmented manually by specifying up to ten pairs of channel ranges
not subject to mask broadening.  A single 82 \kms\ wide channel mask
centered near 0 \kms\ was applied to each IF.

\begin{figure}
\centering
\includegraphics[angle=-90,width=6in]{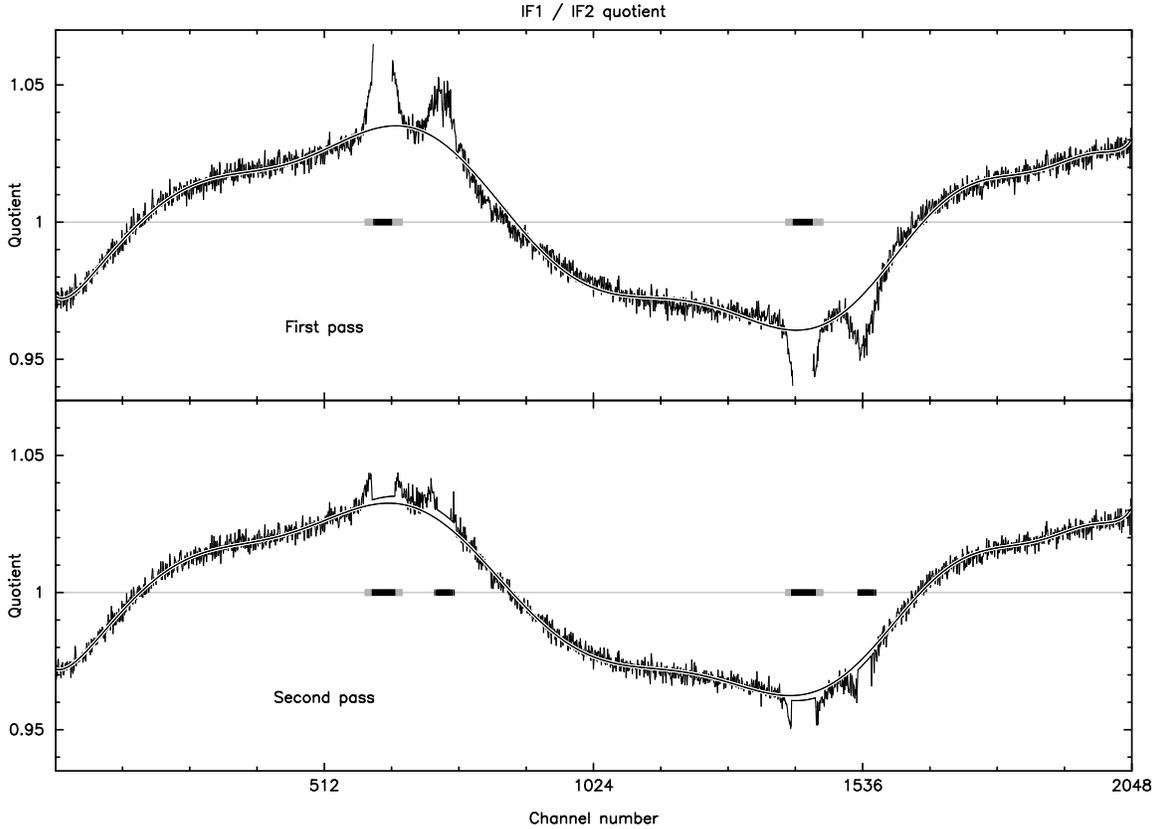}
\caption[]{Average quotient spectra and their polynomial fits for the
  two passes of the bandpass solution described in \S
  \ref{subsubsec:bandpass}.  The  top panel shows the time-averaged and
  masked quotient spectrum, ${\hat q'_1}(\nu)$, overlaid with its
  polynomial fit, $Q'_1(\nu)$.  The user-defined channel mask is
  indicated by the grey solid lines at a quotient value of 1.0, while the 
  black solid lines
  show the derived channel mask.  The polynomial fit in this
  panel is clearly not perfect, caused to a certain extent by the
  parts of the spectral line that are not yet masked.  The lower panel
  shows the second pass in which $Q'_1(\nu)$ is used to refine the
  time masks and produce a new average, ${\hat q''_1}(\nu)$.  The
  second pass fit, $Q''_1(\nu)$, is the bandpass solution.  In this
  panel the polynomial from the first pass is shown in the masked
  channels to indicate how the second-pass fit differs. 
  \label{fig:passes}}
\end{figure}

Broadening of the channel mask is more exaggerated than for the time
mask to avoid removing spectral line wings.  As
they are of scientific interest, it is also important that the wings
not be included in, and therefore potentially removed by, the
polynomial baseline fit which is applied in the next step.  Channel
masking results in gaps in ${\hat q'_1}(\nu)$ which tend to coincide
with channels of particular interest.  At low velocities the \HI\ line
occupies the whole scan; baseline information in these channels is
effectively lost and can only be estimated by interpolation of
neighbouring channels.  In practice ${\hat q'_1}(\nu)$ derived from
observed data, and shown in Figure~\ref{fig:passes}, deviates by a few
percent from unity.  Consequently we found that a robust polynomial
fit of degree 15 was required to interpolate across the masked
channels and fit the baseline accurately.  Robust polynomial fitting
for GASS is an iterative process whereby the polynomial is fit to the
unmasked channels and points outside $3 \times$ the median absolute
deviation from the median are excluded from the second and final
iteration. A second approximation, $Q'_1(\nu)$, to the bandpass
solution is thus obtained, thereby completing the first pass.  

We found that the high order polynomial fit to ${\hat q'_1}(\nu)$
works well for most of the sky, but for regions where the line
emission is spectrally broad the fit is poorly constrained under the
line.  This is mainly an issue towards the Galactic plane and results
in errors at the 3-5\% level as discussed in \S \ref{sec:baseline}.
Lower order polynomials were also tested but we found that they were
not able to fit the substructure of the quotient spectrum.

The quotients and fits for an example scan in the direction of the
Magellanic Stream are shown in Figure~\ref{fig:passes}.  This scan is
a particularly difficult case because the Galactic and Magellanic
Stream emission are spectrally near to each other with very little
spectral baseline between.  The top panel of Figure \ref{fig:passes}
shows the results of the first pass, where the masked, time-averaged
quotient spectrum, ${\hat q'_1}(\nu)$, is plotted together with its
polynomial fit, $Q'_1(\nu)$.  The plotted quotient spectrum
demonstrates clearly the need for the 15th degree polynomial fit.  The
polynomial fit in this panel is, however, not perfect, caused to a
large extent by the parts of the spectral line that are not yet
masked.

The lower panel in Figure~\ref{fig:passes} shows the results of a
second pass through the data, which improves the masking of emission.
The second pass essentially repeats the first pass except that $Q_1$
is replaced with $Q'_1(\nu)$, which provides a channel-specific value.
$D_1$ is recomputed accordingly but the discriminant for time masking
is set at $2 D_1$ (rather than $3 D_1$) because $Q'_1(\nu)$ is a more
reliable estimate of the base-level than $Q_1$.  The new time average
of the more effectively masked quotient spectra, $q''_1(\nu)$, is
shown in the lower panel of Figure~\ref{fig:passes}.  This spectrum is
once again fit with a robust polynomial of degree 15 to give the final
bandpass solution, $Q''_1(\nu)$, as shown in Figure~\ref{fig:passes}.
Here we can see that the more extensive masking results in a better
fit to the off-line portions of the spectrum, although deviations
remain at the $\sim 0.1$\% level.

The bandpass corrected spectra are thus calculated from $Q''_1(\nu)$ as:
\begin{eqnarray}
  S'_{i_1}(\nu) & = & {\rm norm}(q_{i_1}(\nu) /Q''_1(\nu))
                      T_{i_1} - \overline T_1, \nonumber \\
  S'_{i_2}(\nu) & = & {\rm norm}(q_{i_2}(\nu) \times Q''_1(\nu))
                      T_{i_2} - \overline T_2, \nonumber
\end{eqnarray}
where ${\rm norm}()$ indicates normalization to unit value, $T_{i_1}$
and $T_{i_2}$ are the system temperatures for the two spectra, and
${\overline T_1}$ and ${\overline T_2}$ are the median values of
$T_{sys}$ for the two frequency-switched pairs evaluated over all $i$
in the scan.

\subsubsection{RFI Flagging}
Occasional, weak, narrow-line RFI appears at a fixed topocentric
frequency in much or all of some scans.  Once Doppler-shifted to the
LSRK, such RFI produces low-level features that appear to move across
the sky in successive velocity channels.  Because of
frequency-switching, such features may be negative as well as
positive.

Narrow-line RFI appears in only one or two channels, but may cause
ringing in adjacent channels.  The first step in flagging RFI is to 
compute, for each channel, the time-average value of $S'_{i_1}(\nu)$ 
for all $i$ in the scan.  From this, the running mean computed over 21 channels 
honouring the channel mask determined previously, is subtracted.
The result is converted to a mask with values -1, 0, or +1,
depending on whether the channel value is outside $6 \times$ the median
absolute deviation; the magnitude of the departure from zero is not
considered, only the sign.

The mask is then scanned for non-zero channels that should be
preserved, the remainder being flagged as likely RFI.  Consecutive 
non-zero channels of the same sign are preserved if there are more 
than 2 of them in the sequence, the {\em width test}, or if they are close 
to such a sequence of the same sign.  Here, ``close'' means that there 
is no intervening value of the opposite sign, and there is no intervening 
sequence of consecutive zeroes of length greater than 5 channels.  This 
{\em closeness test} is intended to protect outliers in the wings of 
real emission lines.  This algorithm proved effective in removing the 
majority of the RFI that appeared  in spectral datacubes.

\subsubsection{Post-processing}
Low-level baseline residuals may remain in each spectrum at this point
because the original bandpass solution was computed as a time-average
over the whole scan.  This residue was removed separately for each
spectrum by subtracting a 10-th order polynomial fit from the
spectrum for high Galactic latitude areas ($b>10\arcdeg$) and a simple
median level fit from areas within 10 degrees of the Galactic plane.
The algorithm was as described previously, using a mask derived
separately for each spectrum, including the user-defined mask as
before.  We discuss baseline quality in \S~\ref{subsec:quality}.

Finally, the spectra were Doppler-shifted to transform them to the LSRK
velocity frame for gridding and analysis.  The scheme used for Doppler
correction was to rescale the reference frequency and channel spacing
by the Doppler factor, and then Fourier-shift the spectrum, usually by
less than one channel, so that the reference frequency was an integer
factor of the original channel spacing (3.90625 kHz).  This scheme is
employed by {\em Gridzilla} (see below) and allows the combination of
spectra taken months or years apart without requiring interpolation of
the frequency axis.

\subsubsection{Brightness Calibration}

First order brightness calibration was applied on-line through
injection of noise from a diode switched with a frequency of 500 Hz.
Average on-line system temperatures for each beam and polarization
were recorded along with each 5~s spectrum.  Typical system temperature
values are between 21 - 23 K.

Before imaging, spectra are converted to beam averaged brightness temperature,
$T_B$, from observations of the IAU standard line calibration
regions S6, S8, and S9 \citep{williams73,kalberla82a}.  One of the
three standard line regions was observed each day.  Observations were
conducted by placing each of the thirteen beams on the region in turn.
The peak of the observed line was used to calculate calibration
scaling factors for each beam and each polarization assuming peak
brightness temperatures of $T_B = 83$ K for S9, $T_B=53$ K for S6 and
$T_B = 76$ K for S8 valid for the Parkes beam
\citep{kalberla82a,bruens05}.  One set of calibration factors was
calculated for each observing session and showed rms variations
between observing sessions of $1-2$\% on most beams.  The notable
exceptions are the second polarization on beams 10 and 12, which were
known to have unstable low noise amplifiers.  The calibration factors
for these beams varied by as much as 6\% over the 21 months of the
survey.  The overall effect on the data was found to be minimal and
these beams were included in the final data cubes.  These brightness
temperature calibration factors were applied within {\em Livedata}
following bandpass calibration.

To check that our calibration factors were not affected by the
observing strategy of pointed calibrator observations rather than the
on-the-fly scans that were used for the full survey, we observed S9 in
on-the-fly mode as well.  Calibration factors determined from these
scans were fully consistent with the factors determined from pointed
observations.

\subsubsection{Imaging}

Imaging was performed by {\em Gridzilla}, a statistical gridder
developed for the HIPASS survey and subsequently extended for more
general use with Parkes multibeam and other single-dish data.  The
algorithm is described by Barnes et al.\ (2001)\nocite{barnes01}.
Part of {\em Gridzilla's} later development involved adding full
support for FITS celestial and spectral world coordinate systems
\citep{greisen02, calabretta02, greisen06}.  In particular, we used
the Zenithal Equal-Area projection (ZEA) as the most appropriate
choice for mapping the hemisphere.

For each pixel in the output data cube, {\em Gridzilla} computes a weight 
for each input spectrum based on its
angular distance from the pixel.  For GASS, it then calculated the pixel
value from the spectral values and weights using {\em weighted median
estimation}, which is robust against RFI and other artifacts.  The
weighted median of a set of measurements is the {\em middle-weight
value} - the sum-of-weights of all measurements less than it being equal
to that of all measurements greater; pro rata interpolation being used
to bisect the sum-of-weights if required.

For the weighting function we used natural beam weighting, with the
beam modelled by a 2D Gaussian of FWHM $14\arcmin.4$, combined with an
additional 2D Gaussian of FWHM $14\arcmin$, with a cutoff radius of
$8\arcmin$.  This combination of weighting functions degrades the
angular resolution of the images slightly below the telescope FWHM of
$14\arcmin.4$ but produces smoother and more sensitive images.
Measurements of unresolved sources inserted into the data prior to
gridding show that the resulting resolution is $\sim 16\arcmin$.

%----------------------------------------------------------
\section{Data Products}
\label{sec:products}
%---------------------------------------------------------

The primary data product for this first release is a data cube
($\alpha$, $\delta$, $v$) of the entire survey region and full
velocity range without stray radiation correction.  The data were
gridded in a ZEA projection covering $\delta <1\arcdeg$ with the 
South Celestial Pole at the center of the image.  The velocity range 
covered by the gridded cube is $-400~{\rm  km~s^{-1}} \leq V_{LSR} \leq 500$ \kms.  

\begin{figure}
\centering
\plotone{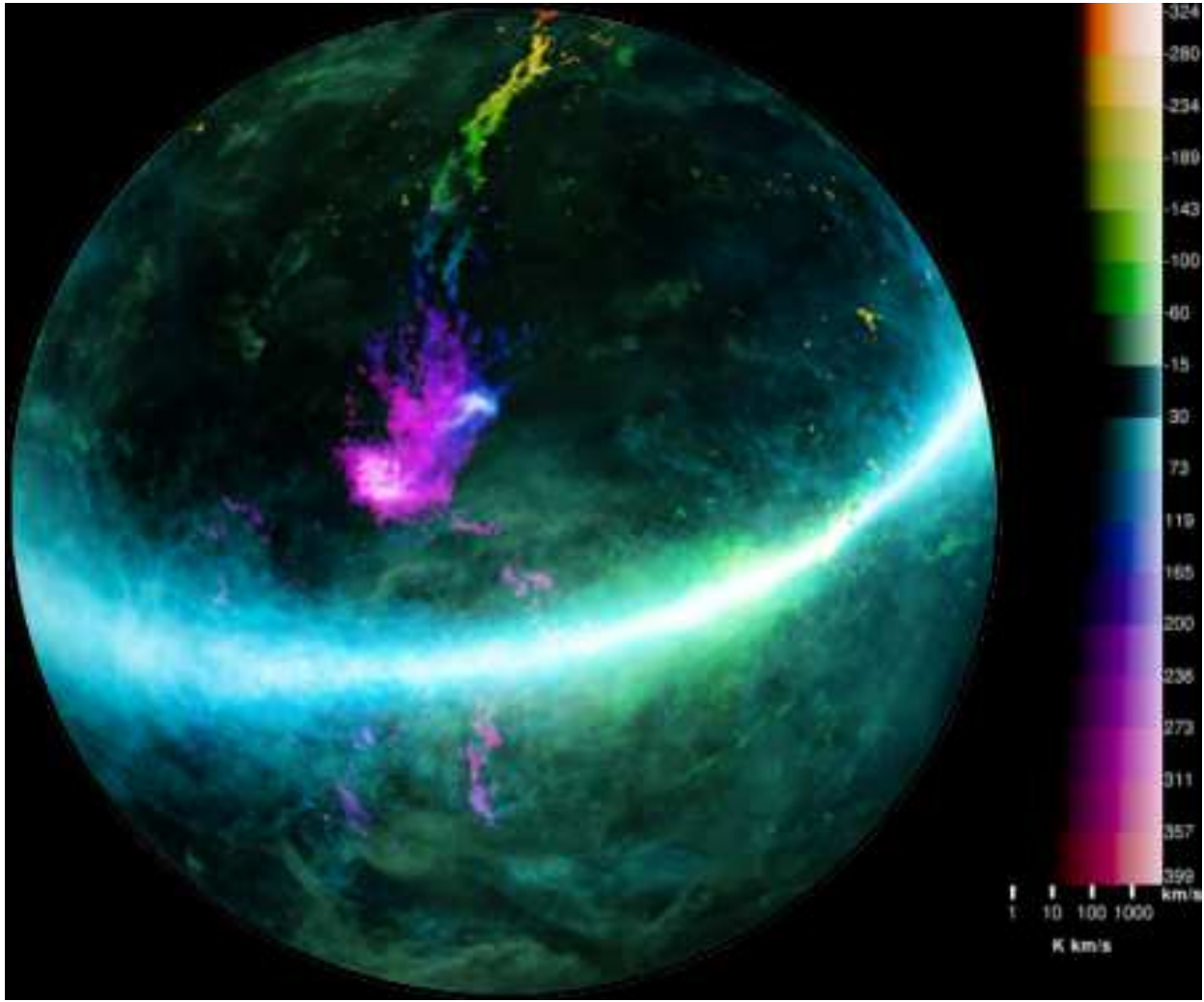}
\caption[]{The entire GASS dataset shown in a ZEA projection
centered on the south celestial pole with 0 hr right ascension at the top
and with RA increasing counter-clockwise.  The colors correspond to integrations
over velocity chunks of $\sim$40 \kms\ as indicated by the bar on the right of 
the image.  The intensity of each color corresponds to the brightness temperature 
integrated over the $\sim$40 \kms\ velocity chunk, and is scaled 
logarithmically as shown by the horizontal extent of the color bar.  Where
emission exists at two different velocities, the intensities are combined
using the ``screen'' algorithm in GIMP as described by \citet{rector07}.  
Some artifacts from scanning were masked by hand.  This image was made 
following the procedure detailed by \citet{rector07}.
\label{fig:fullcolour}}
\end{figure}

A comprehensive view of GASS is given in Figure~\ref{fig:fullcolour},
which is a combination of moment maps created in $\sim 40$ \kms\ intervals
over the full velocity range, color-coded by velocity.  Most of the
key features of the Southern sky are clearly visible in this image
including the Galactic plane, Magellanic Clouds, Magellanic Stream and Leading
Arm, several high velocity cloud complexes, and a few galaxies 
belonging to the Sculptor group.

\begin{figure}
\centering
\includegraphics[angle=-90,width=6in]{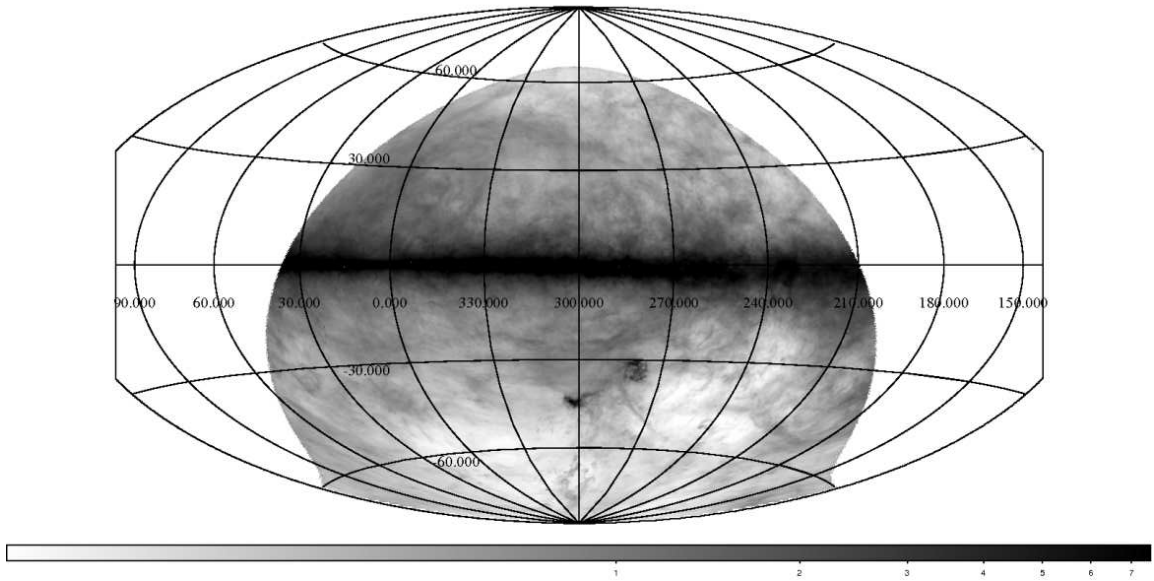}
\caption[]{Total column density image in units of $10^{21}~{\rm
    cm^{-2}}$.  The image is on an Aitoff projection in Galactic
  coordinates centered at $l=300\arcdeg$.  The greyscale is logarithmic and shown in 
the wedge at the bottom.  
\label{fig:NH_ait}}
\end{figure}

An image of the total column density, with an Aitoff projection in
Galactic coordinates, is shown in Figure~\ref{fig:NH_ait}.  The image
has been calculated from the zeroth moment over the velocity range
$-400 \leq V_{LSR} \leq 500$ \kms\ and converted to units of $\times
10^{21}~{\rm cm^{-2}}$ using the usual optically thin scaling factor
$1.8\times 10^{18}~{\rm cm^{-2}~K^{-1}~km^{-1}~s}$.  This image is 
dominated by emission at local velocities.

\begin{figure}
\centering
\plotone{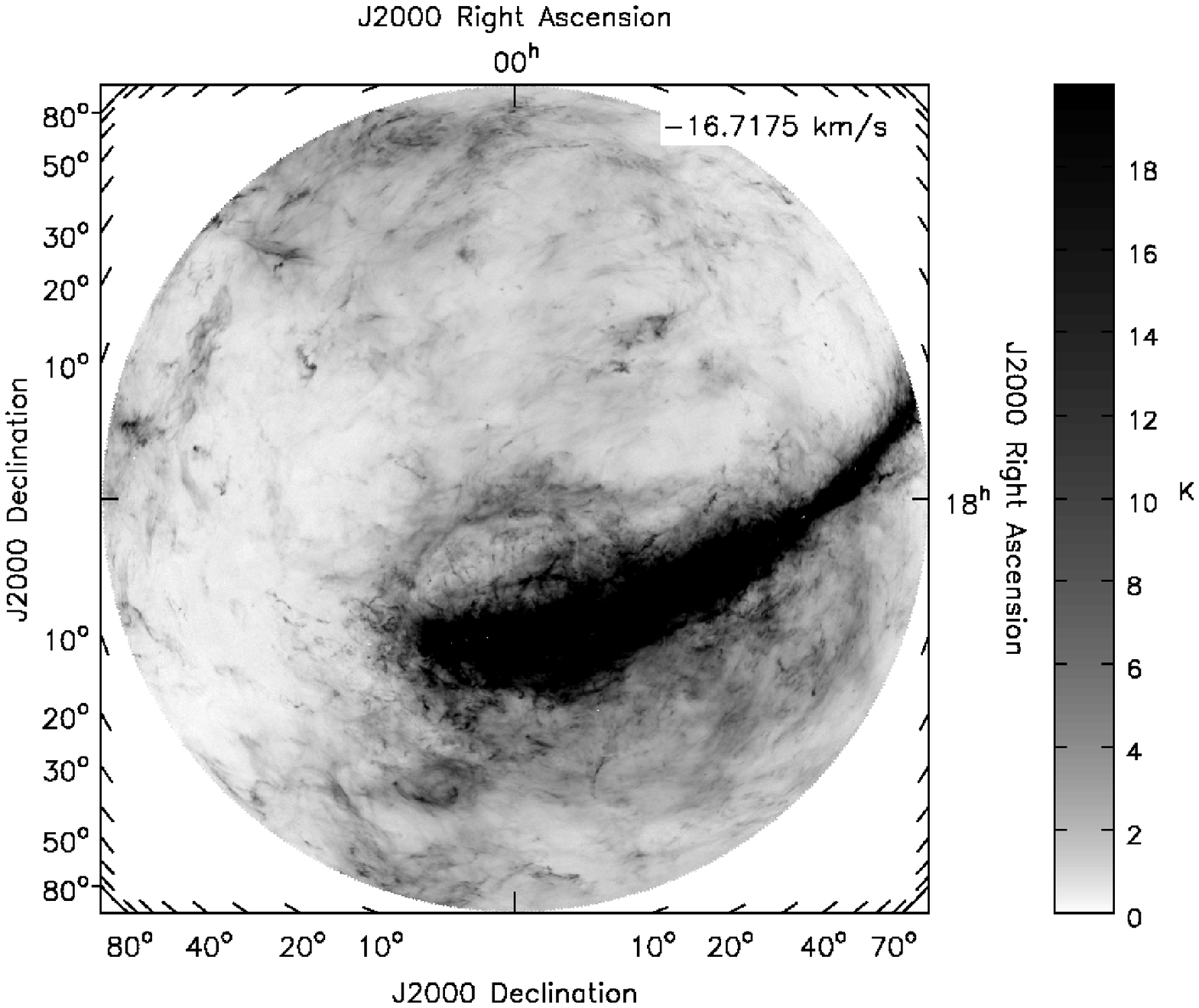}
\caption[]{GASS image at $V_{LSR}=-16.7$ \kms. The greyscale goes from 0 to 20 K
  with a scaling power of $-1$ as shown in the wedge at the right.  
\label{fig:gass-16.7}}
\end{figure}

\begin{figure}
\centering
\plotone{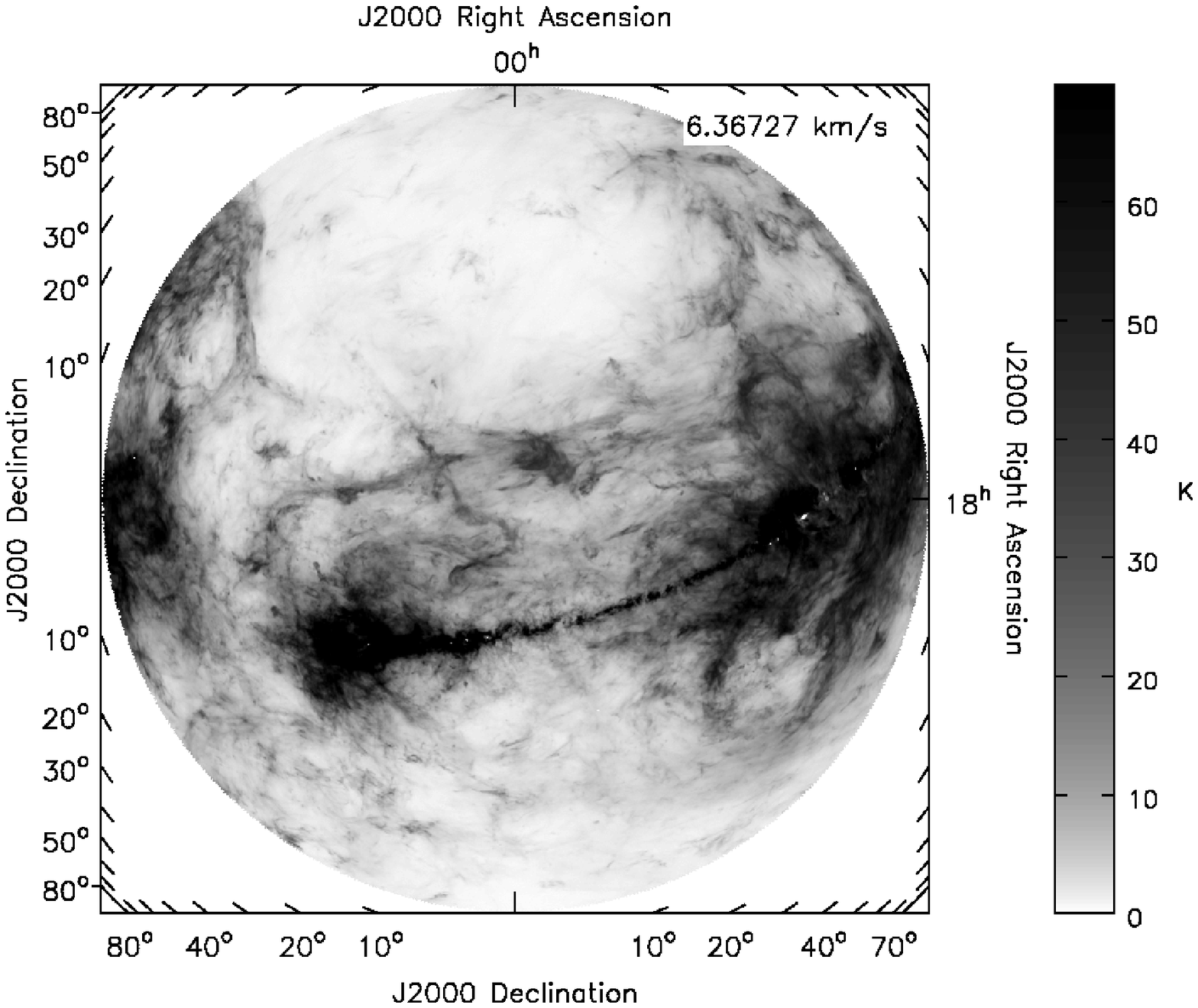}
\caption[]{GASS image at $V_{LSR}=6.36$ \kms. The greyscale goes from 0 to 70 K
  with a scaling power of $-1$ as shown in the wedge at the right. 
\label{fig:gass6.4}}
\end{figure}

\begin{figure}
\centering
\plotone{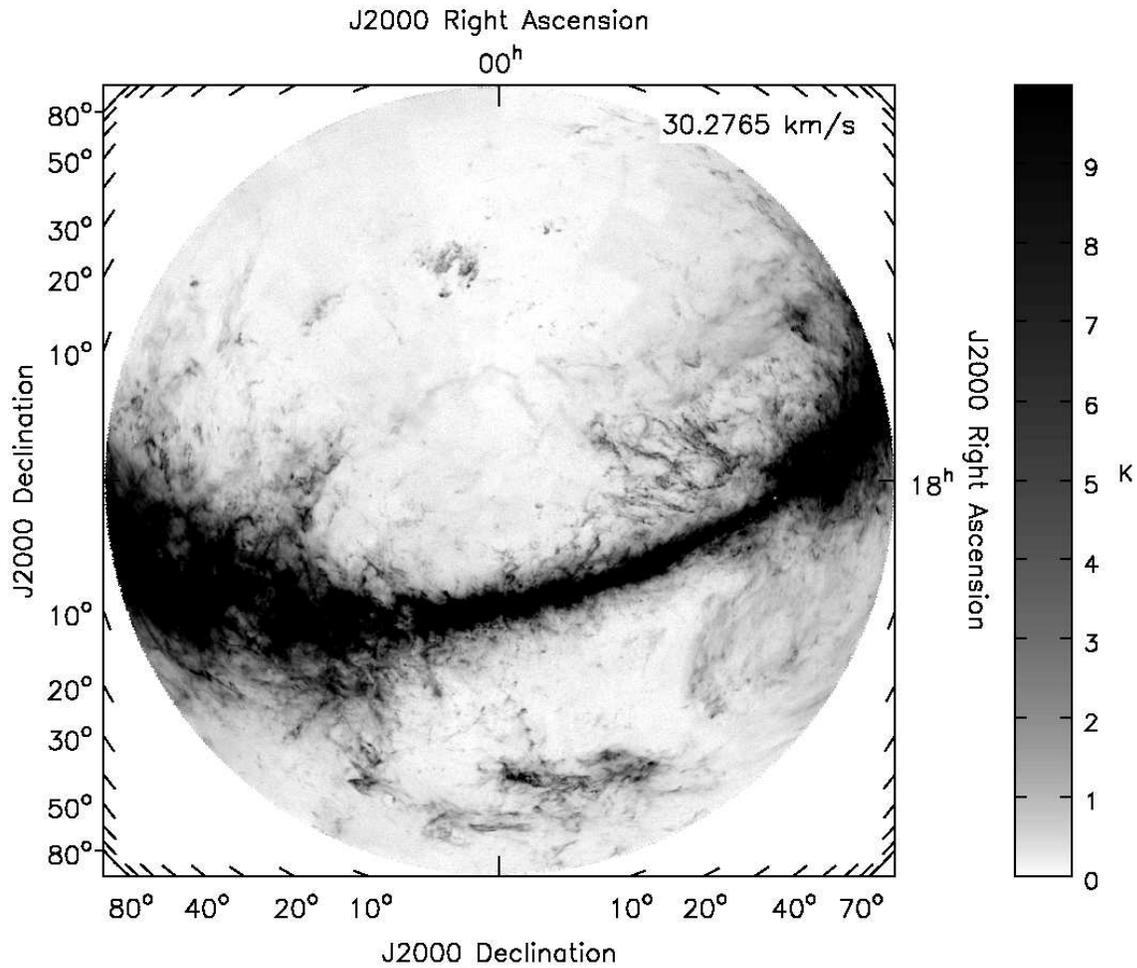}
\caption[]{GASS image at $V_{LSR}=30.3$ \kms. The greyscale goes from 0 to 10 K
  with a scaling power of $-1$ as shown in the wedge at the right.
  The patchiness in the low level emission, particularly near RAs of 0
  h, is evidence of the difference in the stray radiation contribution to
  spatial areas observed in different observing epochs.
\label{fig:gass30.3}}
\end{figure}

\begin{figure}
\centering
\includegraphics[width=4.5in]{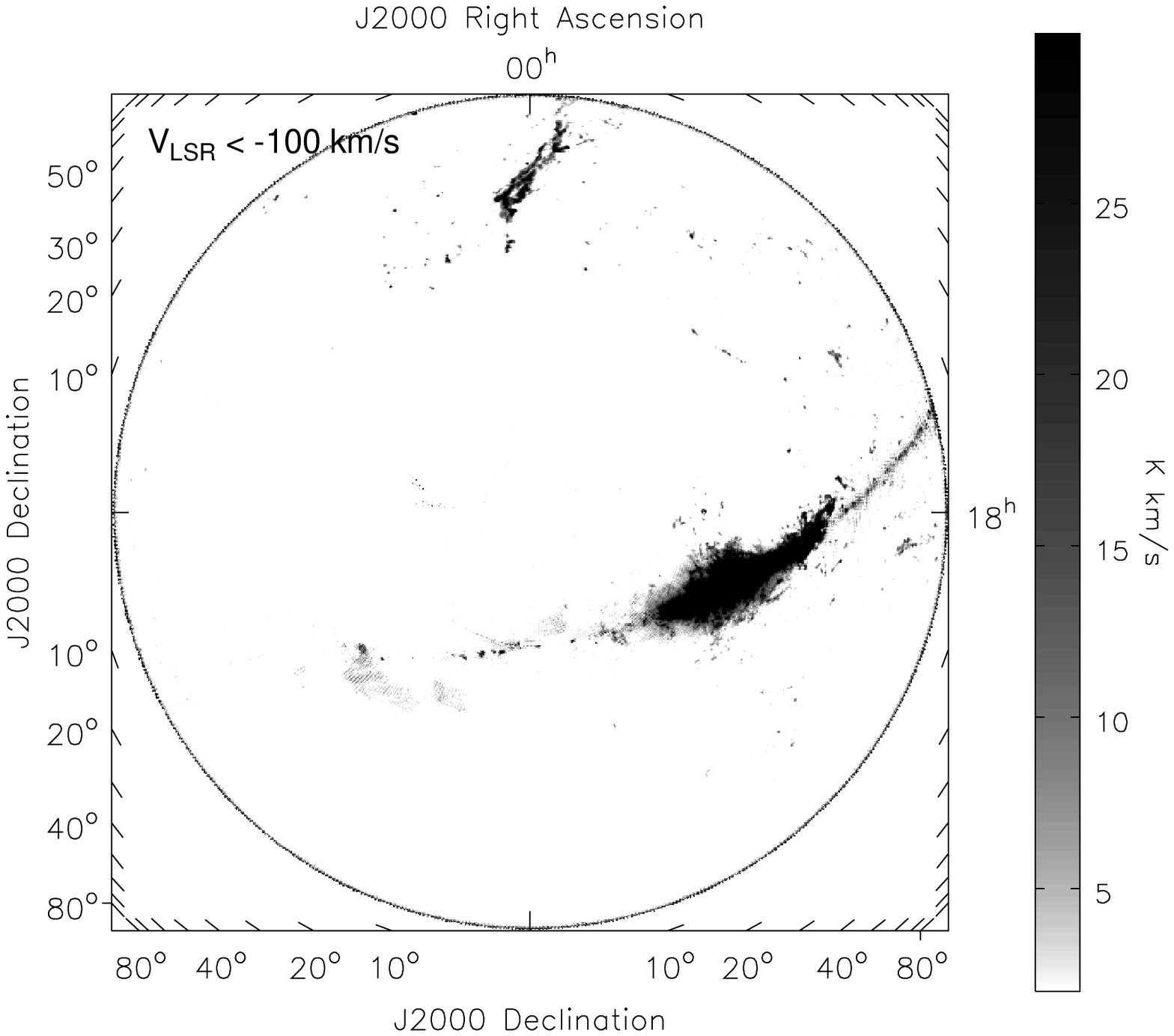}
\includegraphics[width=4.5in]{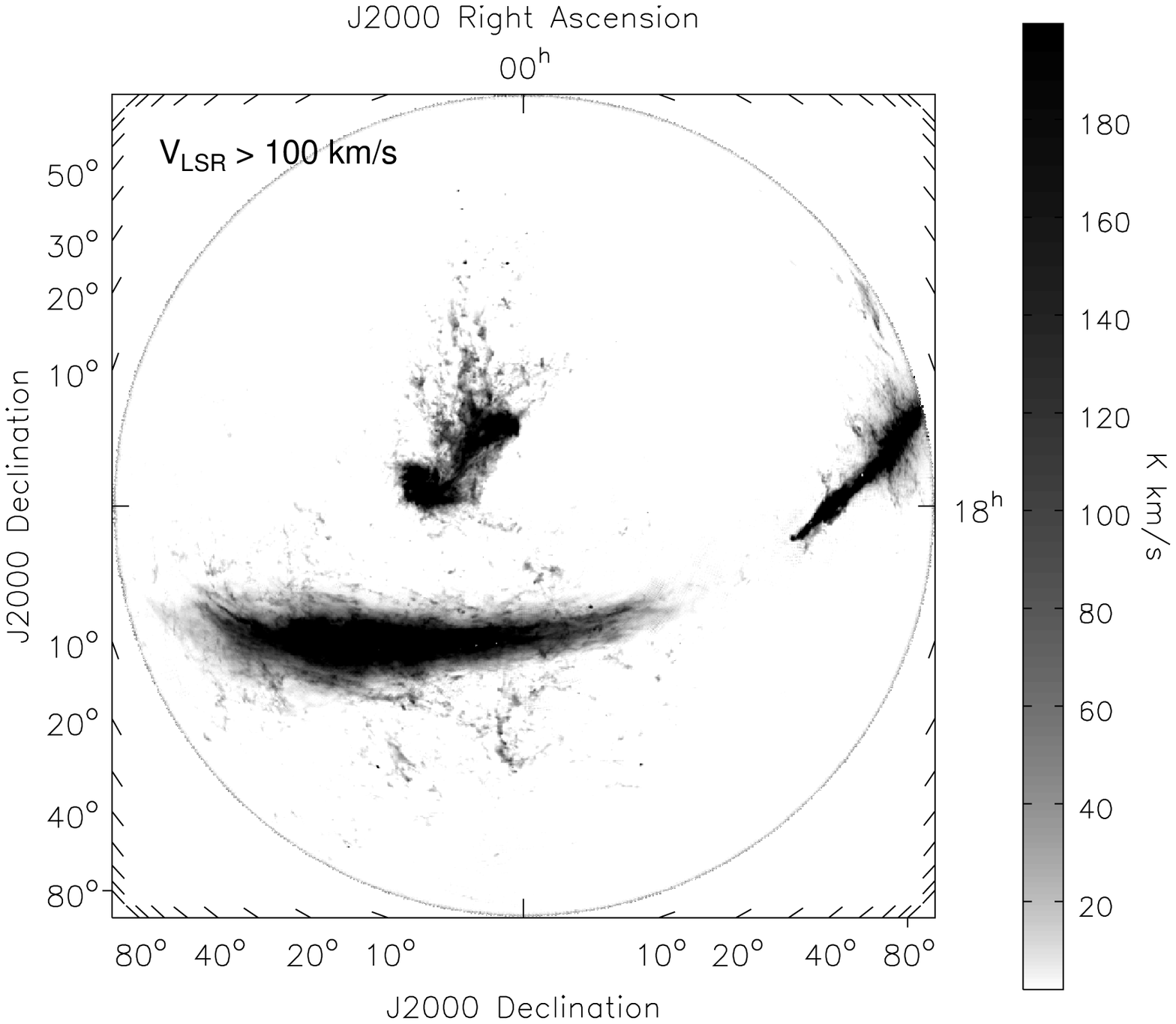}
\caption[]{GASS high velocity sky for negative (top: $V_{LSR}\leq-100$ \kms)
  and positive (bottom: $V_{LSR}\geq 100$ \kms)  velocities.  The greyscales
  use a scaling power of $-1$ and are shown in the accompanying wedges.
  The negative high velocities are dominated by Galactic emission near
  to the Galactic center and a portion of the Magellanic Stream.  The
  positive velocities are dominated by Galactic plane emission and the
  Large and Small Magellanic Clouds.  Small HVCs are visible in both images.
\label{fig:HVC}}
\end{figure}

\begin{figure}
\centering
\includegraphics[width=4.5in]{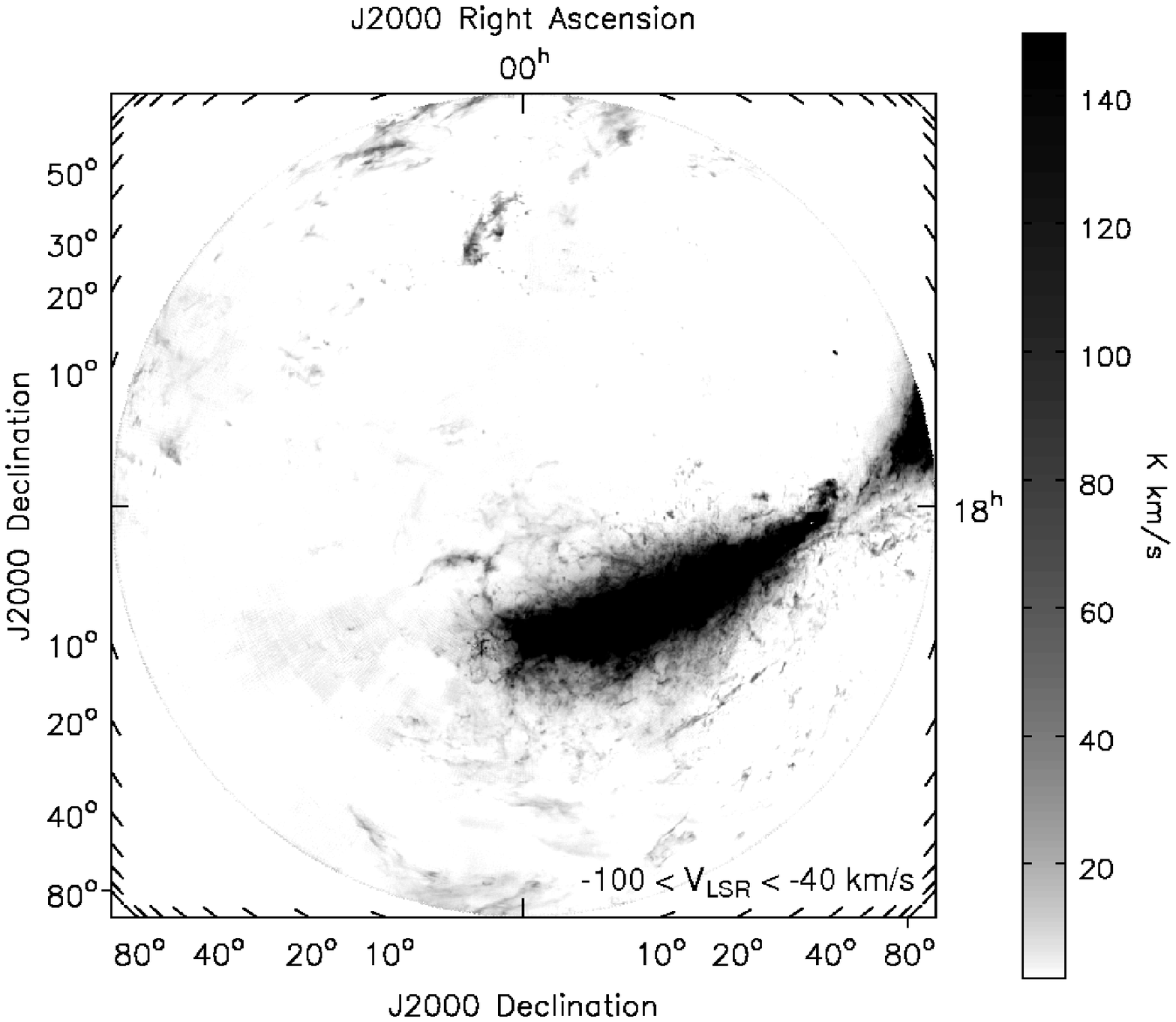}
\includegraphics[width=4.5in]{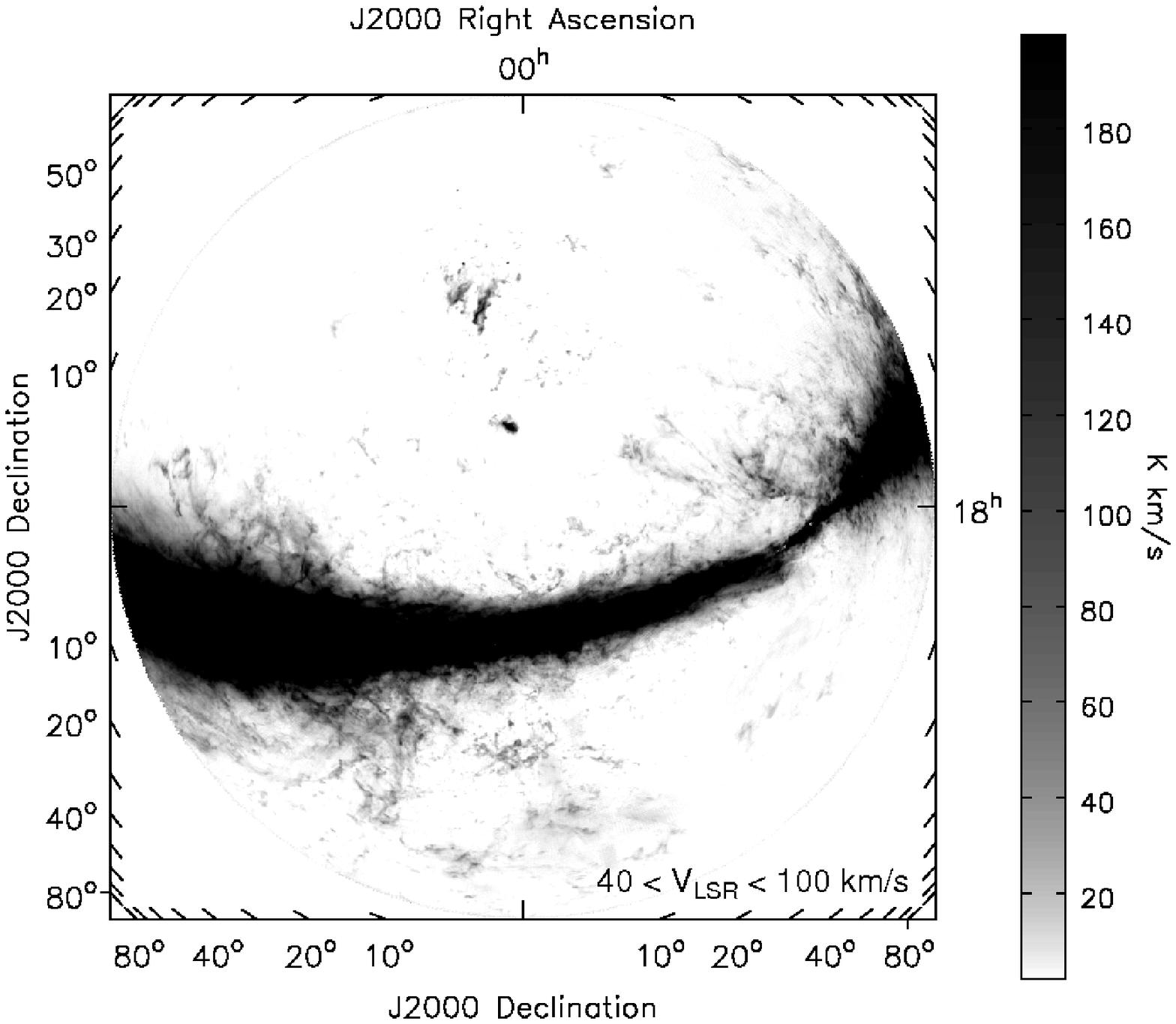}
\caption[]{GASS intermediate velocity sky for negative (top:
  $-100<V_{LSR}<-40$ \kms) and positive (bottom: $40<V_{LSR}<100$ \kms)
  velocities.  The greyscales use a scaling power of $-1$ and are shown
  in the accompanying wedges.  A significant portion of the
  intermediate velocity sky is dominated by Galactic emission, however
  the filamentary nature of the extensions off the Galactic plane are
  noteworthy.  
  \label{fig:IVC}}
\end{figure}

We show three individual velocity channels of low velocity ($|V_{LSR}|
\lesssim 30$ \kms) gas in Figures \ref{fig:gass-16.7} -- \ref{fig:gass30.3}
that highlight the small-scale structure visible in the GASS data.  
Similarly, Figures \ref{fig:HVC} and \ref{fig:IVC} are
images of the high and intermediate velocity sky, where high velocity
gas is defined as $|V_{LSR}|\geq 100$ \kms\ and intermediate velocity
gas is defined as $40 \leq |V_{LSR}| < 100$ \kms.  The negative high
velocities are dominated by Galactic emission near to the Galactic
center and a portion of the Magellanic Stream, whereas the positive
velocities are dominated by Galactic plane emission and the Large and
Small Magellanic Clouds. The intermediate velocity sky is dominated by
Galactic emission, however there are some interesting filamentary
extensions off the Galactic plane.  Small HVCs and IVCs are visible in
both figures.

A summary of the survey specifications is given in
Table \ref{tab:summary}.  Spectra from the survey and information
about other data products available for download can be found at
\break \url{http://www.atnf.csiro.au/research/GASS}.

\subsection{Data Quality}
\label{subsec:quality}

\begin{figure}
\centering
\includegraphics[width=6in]{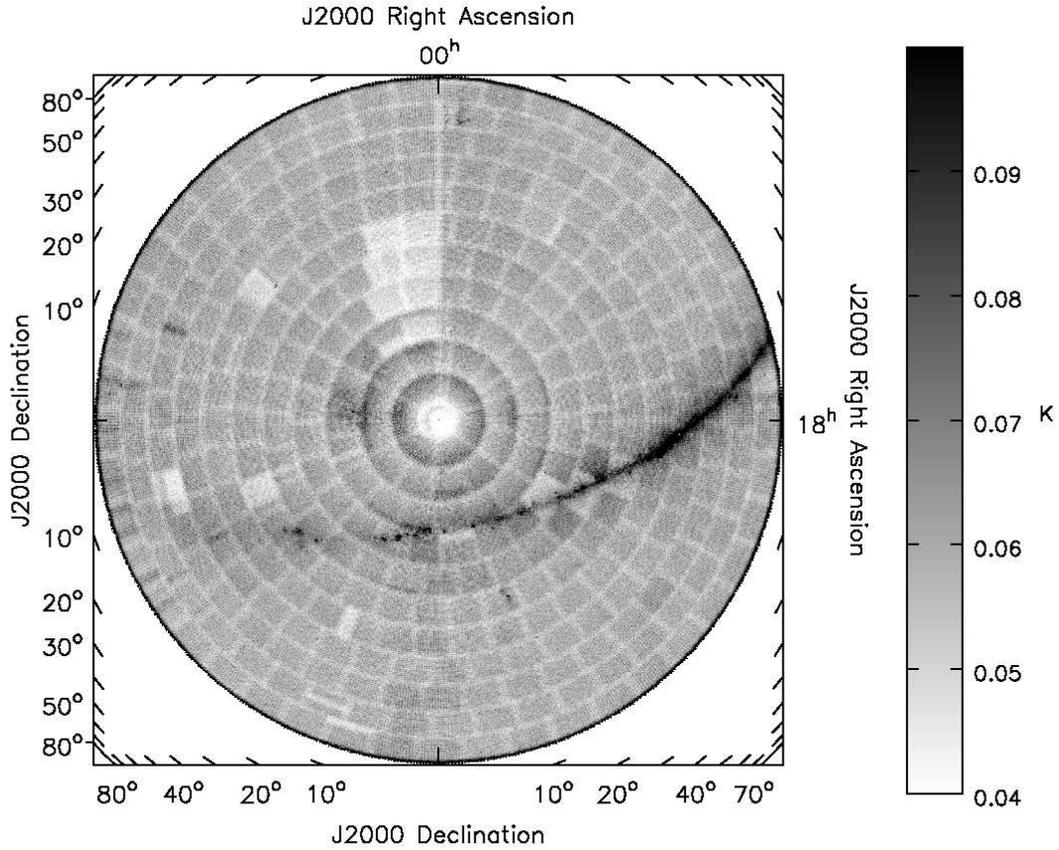}
\caption[]{RMS across the GASS field calculated as the median of the
  rms in nine different blocks of $\sim 21$ emission-free channels.  The greyscale is 
  displayed in the color wedge to the right.  The mode of the rms is 57 mK across
the field, with higher values in the Galactic Plane, towards Cen A and
towards the Magellanic Clouds.
\label{fig:rms}}
\end{figure}

The typical rms per channel in the cube is $\sim 57$ mK.  This is
determined from an image of the rms noise across the full survey
region, as calculated from the median of the noise in nine blocks of
$\sim 20$ line-free channels.  The rms image is shown in
Figure~\ref{fig:rms}.  This approach effectively removes contamination
of the rms measurement from negative residuals (see
\S~\ref{sec:artifacts}) from the Magellanic system and Galactic Center
region in the otherwise line-free channels. The noise is relatively
smooth across the majority of the sky, with a mode of $ 57$ mK.  As
seen in Fig.\ \ref{fig:rms}, the rms around the Galactic plane region
is clearly higher than the rest of the sky. This is due to both strong
continuum sources and the large fraction of the band filled by \HI\
emission in the Galactic plane.  Both effects raise the system
temperature.  Similar increases in rms occur in the region of the LMC.
The rms is lower in the overlap region between scans and also where
scans were repeated, such as the large area near 1 h of right
ascension.  Assuming a rms brightness temperature of $57$ mK, the
$1\sigma$ column density noise in a single velocity channel is
$1\times 10^{17}~{\rm cm^{-2}}$.  For a typical HVC of width 30 \kms,
the $3\sigma$ sensitivity limit is $N_{HI} = 1.6\times10^{18}~{\rm
  cm^{-2}}$.

There are three extended areas where portions of the spectra in the
range velocities $\sim 230 - 320$ \kms\ have increased noise by a
factor of $\sqrt{2}$.  These regions result from two days' worth of RA
scans that were flagged because they exhibited nearly continuous,
broadband RFI of unknown origin.  The flagged scans lie in the
regions: $11^{\rm h} 15^{\rm m} \lesssim \alpha \lesssim 12^{\rm h}
15^{\rm m}$, $-25\arcdeg \lesssim \delta \lesssim 0\arcdeg$; $11^{\rm
  h} \lesssim \alpha \lesssim 13^{\rm h}$, $-65\arcdeg \lesssim \delta
\lesssim -45\arcdeg$; and $13^{\rm h} 45^{\rm m} \lesssim \alpha
\lesssim 15^{\rm h} 15^{\rm m}$, $-78\arcdeg \lesssim \delta \lesssim
-70\arcdeg$.  These areas are marginally visible in
Figure~\ref{fig:rms} as regions of increased noise.

\subsubsection{Stray Radiation}

A fundamental limitation of GASS data in the present release is
stray radiation: \HI\ emission that enters the receiver from the sidelobes 
of the telescope rather than through the main beam.  \citet{kalberla80} has 
shown that stray radiation can make a significant contribution (15 -- 50\% of 
the profile area) to observed Galactic \HI\ emission profiles, mostly at 
high Galactic latitudes where emission in the primary beam is weak.  
The LAB survey has been corrected for stray radiation, and the second 
GASS data release will be corrected as well, but  at this 
stage we have simply estimated  the total amount stray radiation in GASS 
for various regions of the sky.
This was done by convolving the GASS and LAB surveys to  $2\arcdeg$ 
angular resolution, computing the total column density over their common
velocity range ($-400~{\rm km~s^{-1}} \leq V_{LSR} \leq 400$ \kms), and 
interpreting the difference as stray radiation in GASS.
Figure~\ref{fig:stray} shows the convolved GASS column density, the estimated 
stray column, and the stray fraction.   For most lines of
sight the fraction of the total column density believed to arise from 
stray radiation is between 5 and 15\%, with some
particularly low column density regions showing fractions as high as
35\%.

\begin{figure}
\centering
\includegraphics[width=6.5in]{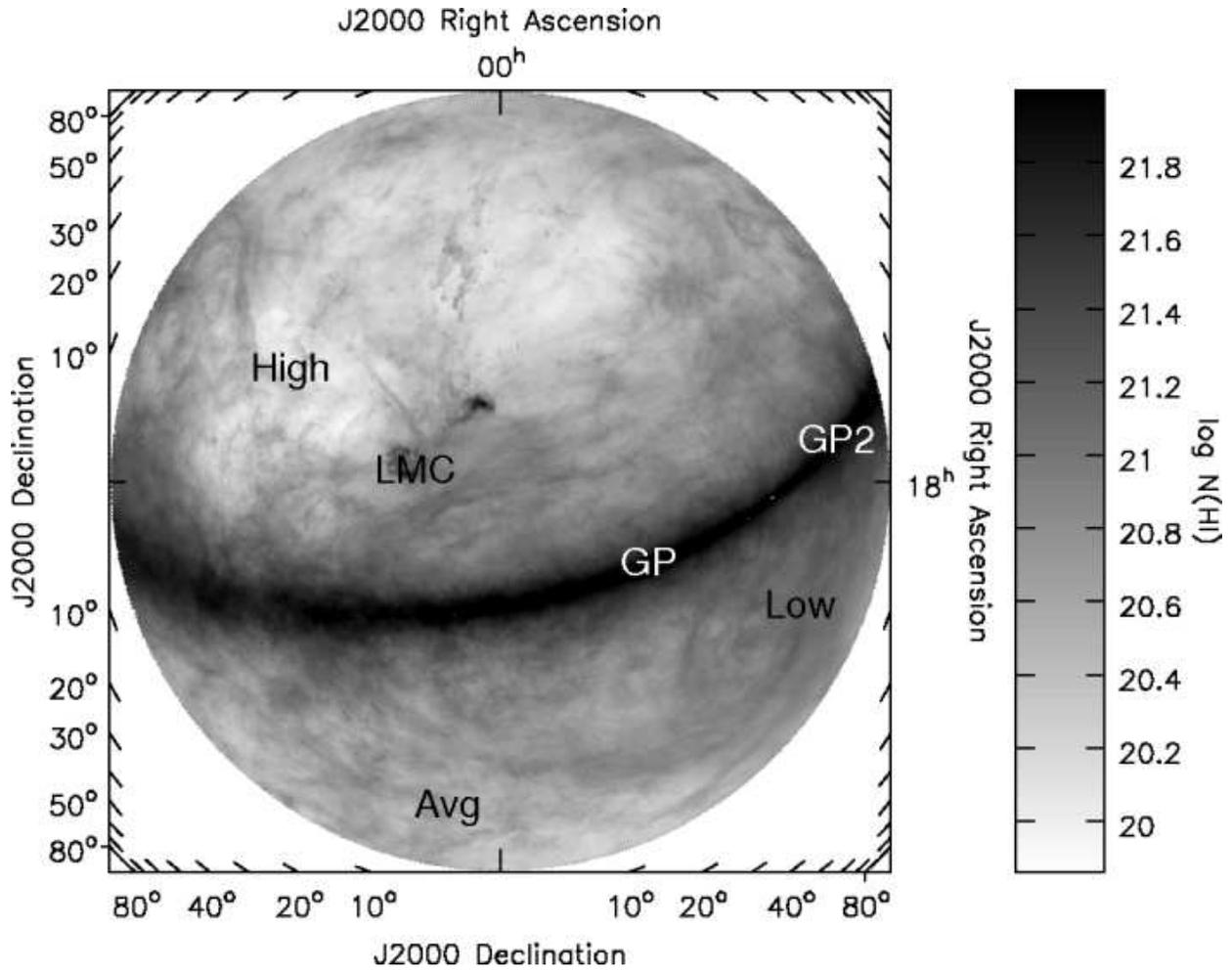}
\caption[]{Annotated total column density image for GASS with the areas of low,
  average and high stray fraction used in  Figure~\ref{fig:spec1} marked together 
  with the Galactic Plane and LMC areas referred to in Figure~\ref{fig:spec3}.  The
  greyscale is $\log{N_{HI}}$ from 19.8605 to 22 in units of cm$^{-2}$.
\label{fig:column_annotate}}
\end{figure}

\begin{figure}
\centering
\includegraphics[width=3.3in]{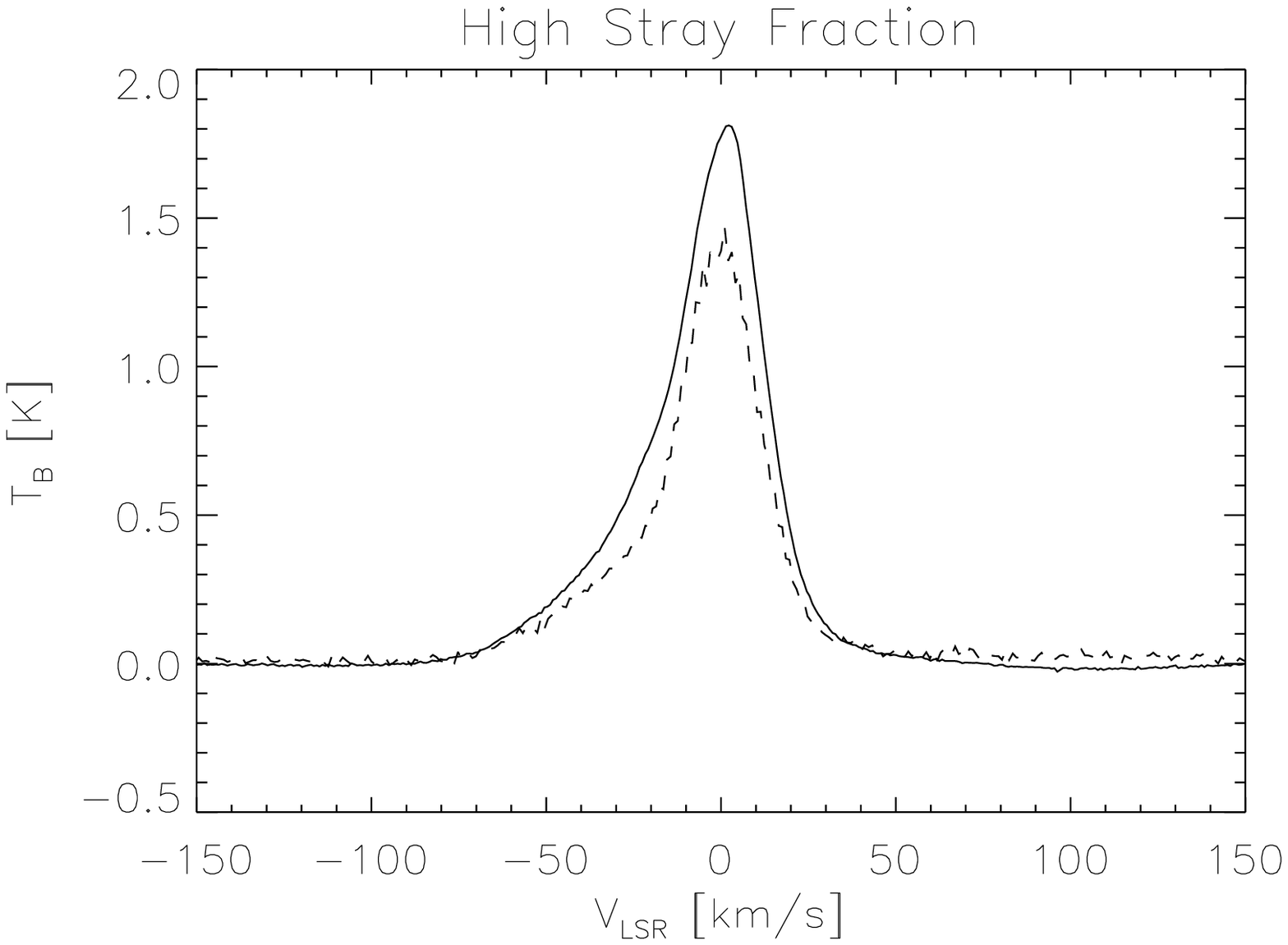}
\includegraphics[width=3.3in]{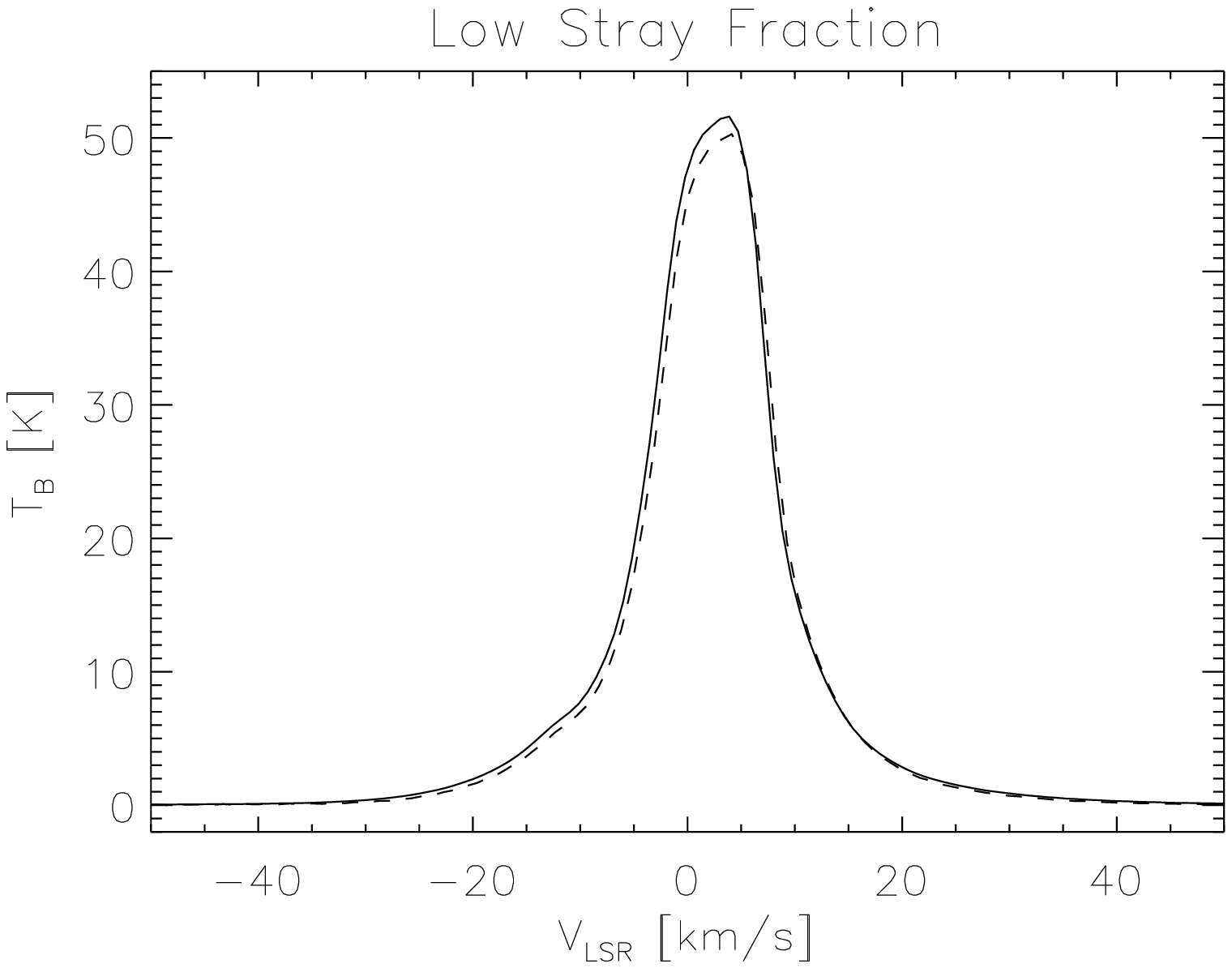}
\includegraphics[width=3.3in]{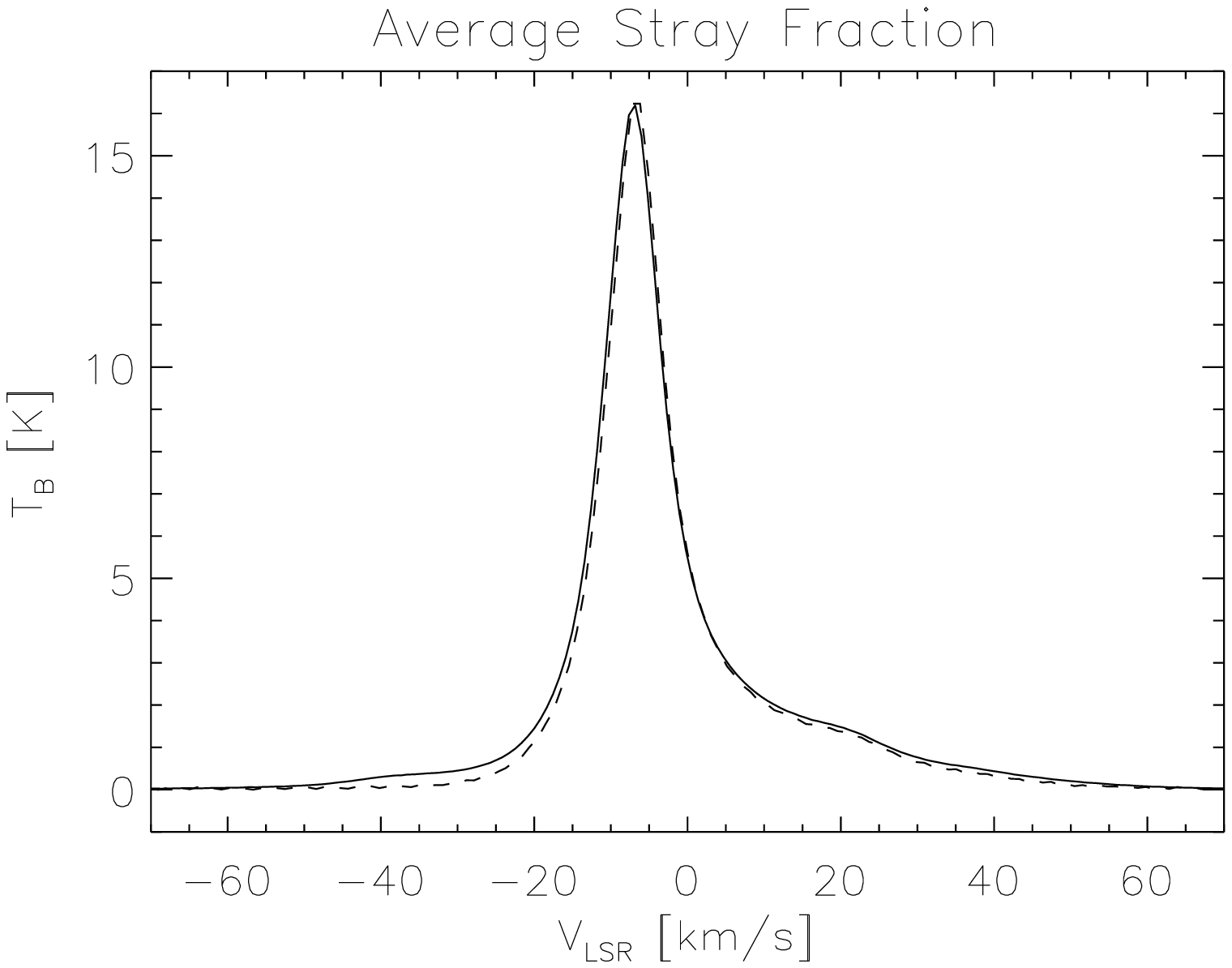}
\caption[]{Comparison of spectra from GASS (solid line) and LAB
  (dashed line) towards the areas marked in
  Figure~\ref{fig:column_annotate}.  These spectra, smoothed over 
$5\arcdeg \times 5\arcdeg$ region, show the effects of
  stray radiation, which produces slightly higher spectra in GASS than
  LAB and line wings.
\label{fig:spec1}}
\end{figure}

To demonstrate the spectral behaviour of the stray radiation component
we compared LAB and GASS spectra towards several regions
representative of high stray radiation fraction, low stray fraction,
and average stray fraction, as calculated above.  The regions are
marked on Figure~\ref{fig:column_annotate} and the spectra are shown
in Figure \ref{fig:spec1}.  These spectra are averaged over $5\times
5$ degrees in both surveys to minimize any resolution-dependent
effects.  For an area with average stray fraction it is clear that the
GASS spectrum reproduces the stray radiation corrected LAB spectrum
very well, with small departures in the line wings where the total
intensity is small.  The high stray fraction spectrum was extracted
towards a region where the fraction of the total column density
attributed to stray radiation was on the order of 30\%.  In this area
the GASS spectrum has a very low peak brightness temperature of $\sim
1.9$ K, and yet the LAB spectrum peak is even lower at $\sim 1.4$ K.
The GASS spectrum also shows significant effects of stray radiation in
the line wings.

\subsubsection{Baseline Quality}
\label{sec:baseline}
Although GASS spectral baselines are generally very good, there are
residual ripples in the final baselines with maximum peak-to-peak
variations of $\sim 50$ mK.  It was the presence of much larger ($\sim
100 - 200$ mK) bandpass amplitude ripples and striping in test images
that necessitated the use of high-order polynomials in the bandpass
correction algorithm as described in \S \ref{sec:details}.  These were
effectively removed by the post-bandpass polynomial fits applied to
scans away from the Galactic Plane.  To ensure that the high-order
polynomials had not subtracted real emission we compared spectra
obtained from GASS to spectra obtained from the LAB survey. These
showed generally very good agreement at the $\sim 50-70$ mK
level. Several examples are shown in Figure~\ref{fig:spec_comparison},
demonstrating the baseline quality of GASS.

\begin{figure}
\centering
\includegraphics[width=3.3in]{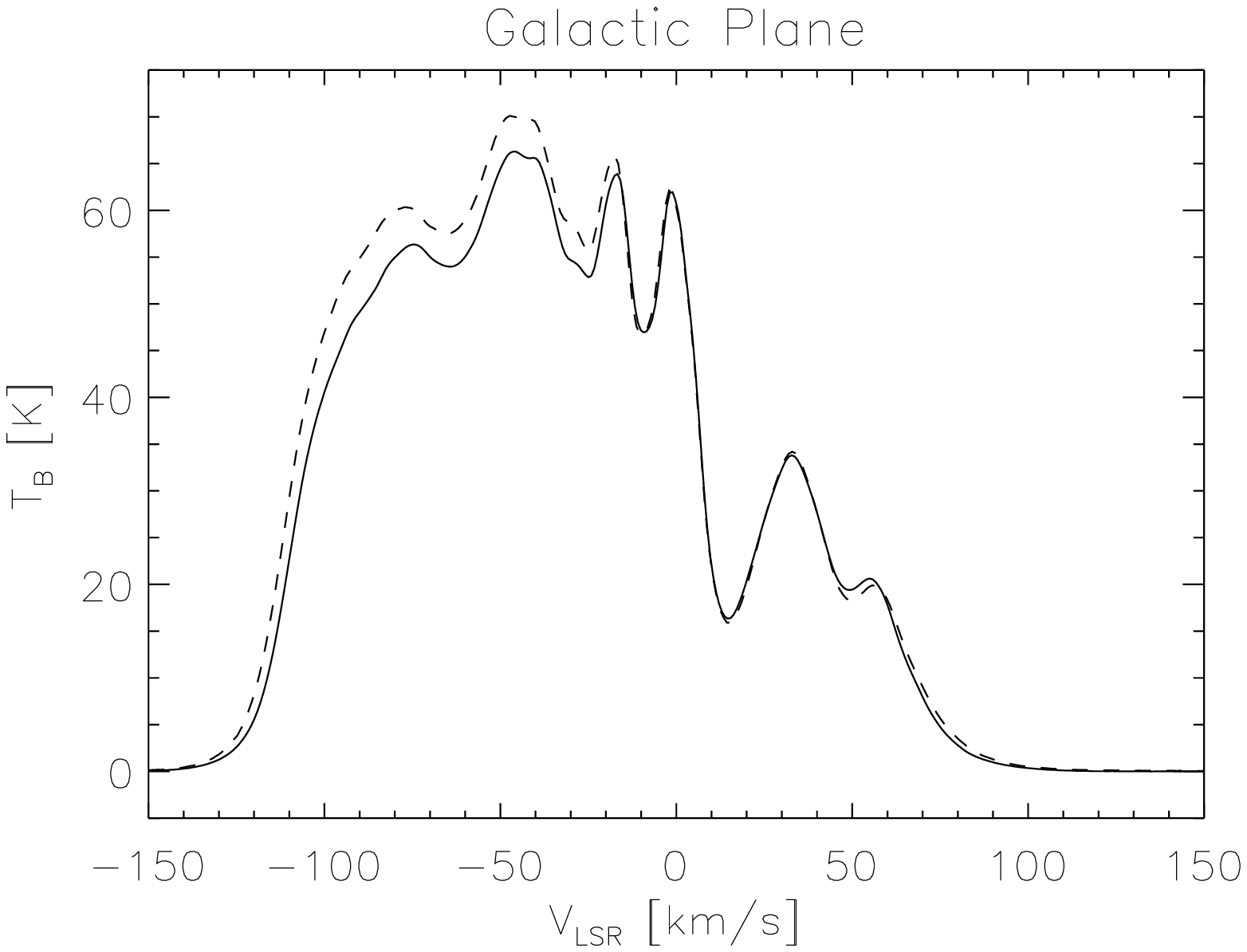}
\includegraphics[width=3.3in]{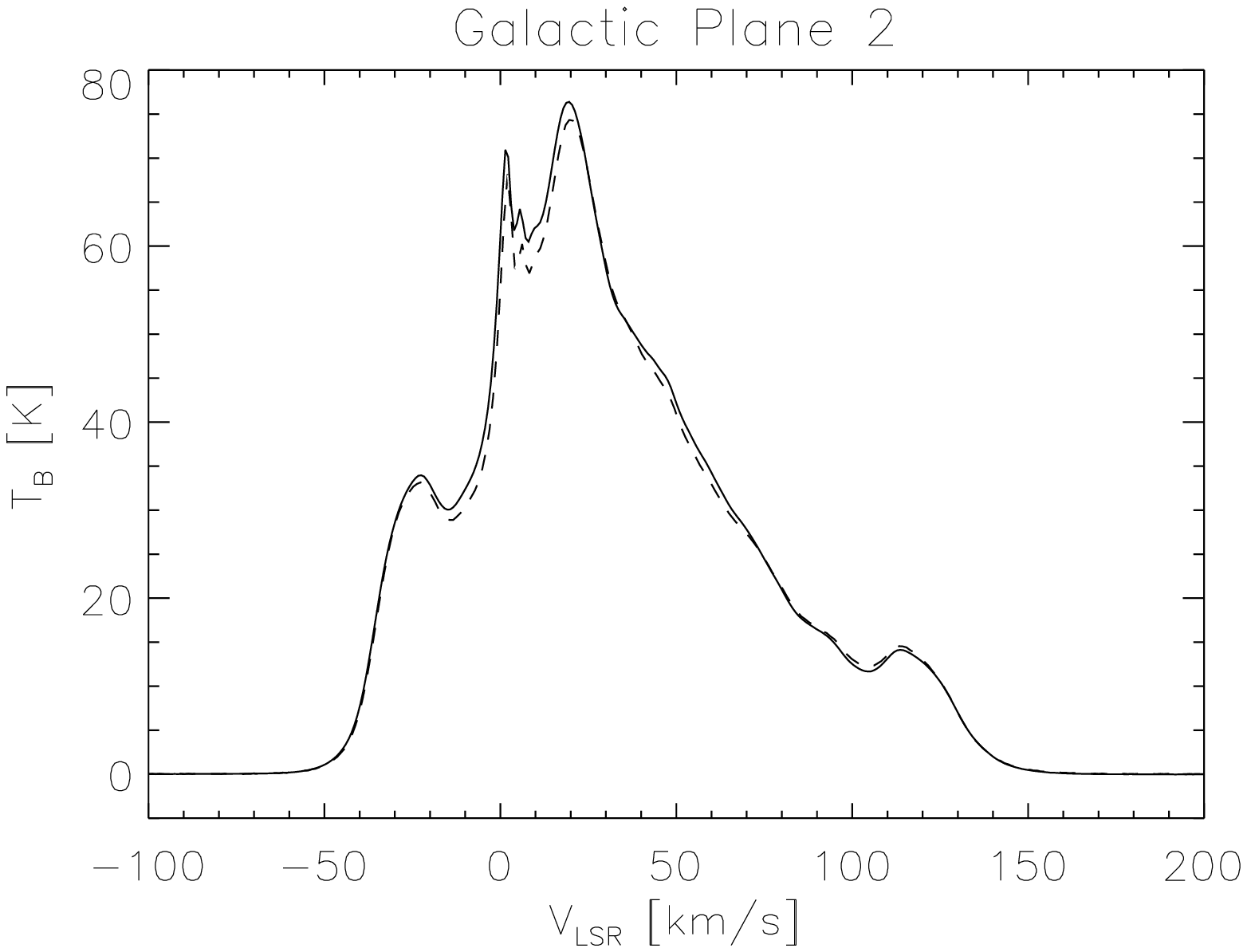}
\includegraphics[width=3.3in]{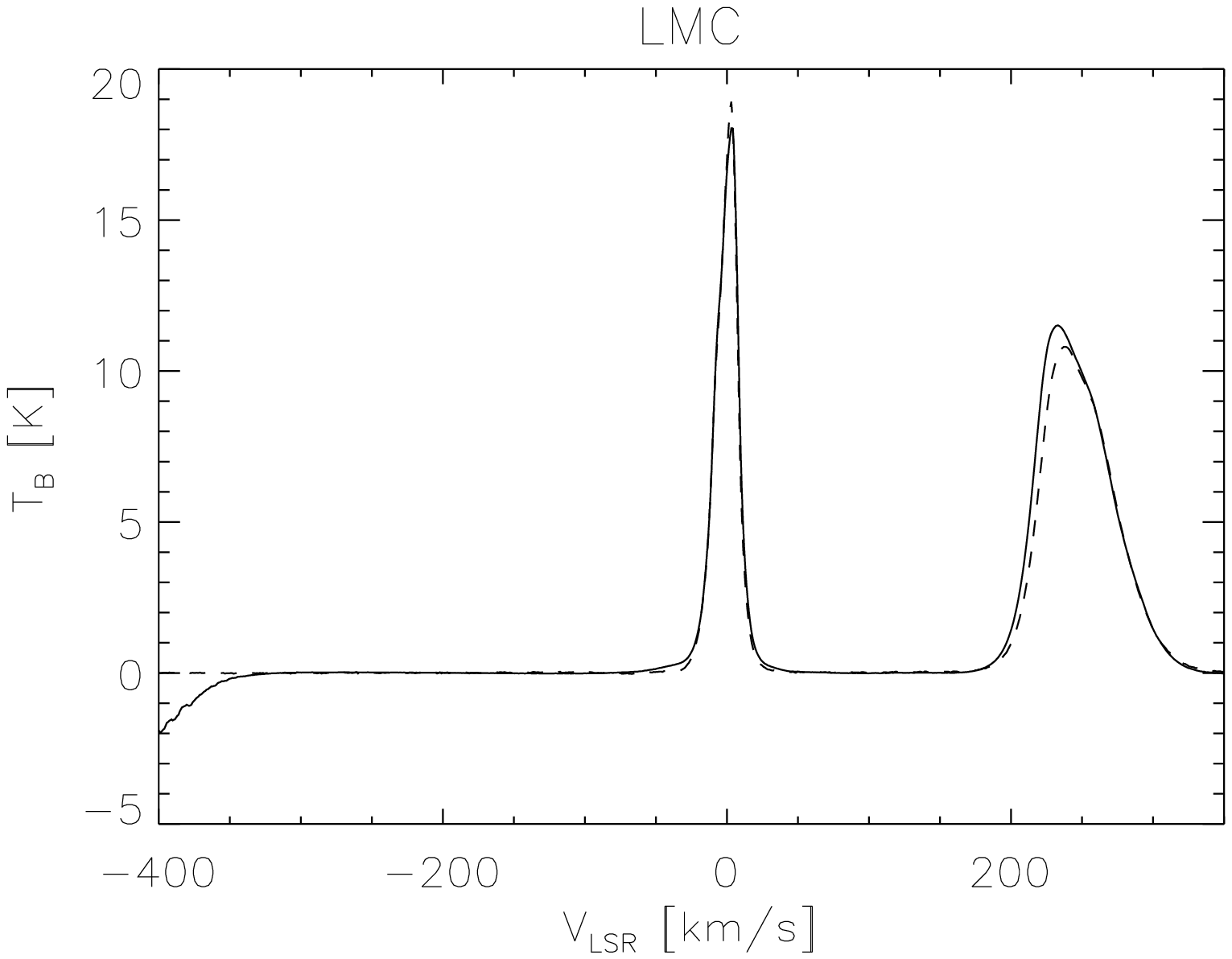}
\caption[]{Comparison of spectra from GASS (solid line) and LAB
  (dashed line) towards the areas marked in
  Figure~\ref{fig:column_annotate} as GP (top), GP2 (middle), and LMC
  (bottom).  The spectra at the top and in the middle show the
  limitations in GASS bandpass solutions in the Galactic plane, which
  result in spectral distortions relative to LAB.  The bottom spectrum
  shows the negative image at $V_{LSR}=-400$ \kms\ from the LMC
  (located at $V_{LSR}=250$ \kms) caused by in-band
  frequency-switching.  These spectra are smoothed over a $5\arcdeg
  \times 5\arcdeg$ region.
  \label{fig:spec3}}
\end{figure}

Baseline quality deteriorates in some regions toward the Galactic
Plane and the Magellanic Clouds.  The bandpass correction method
described in \S \ref{subsubsec:bandpass} is limited in cases where the
\HI\ emission in the quotient spectrum fills $\sim 50$\% of the band.
In these cases the polynomial fit to the quotient spectrum is poorly
constrained and therefore the bandpass under the line is not well
determined.  This results in some small, $\sim 3-5$\%, baseline errors
where the measured line temperatures can appear either too large or
too small relative to spectra from LAB.  Examples are shown in
Figure~\ref{fig:spec3}, where the GASS spectra show higher and lower
peak values than LAB.

\subsubsection{Other Artifacts}
\label{sec:artifacts}

As discussed in \S~\ref{subsec:obs}, the in-band frequency-switching
will produce a negative copy of a spectral line displaced by $\pm$660
\kms.  As a practical matter, most of these artifacts lie outside the
final data cube, but there are some notable exceptions for features at
$V_{LSR}>260$ \kms\ (with the inverted feature appearing at negative
velocities) and $V_{LSR}<-160$ \kms\ (at positive velocities).  This is
particularly evident towards the Galactic Center, the Magellanic
Clouds (apparent at extreme negative velocities) and the Northern tip
of the Magellanic Stream (apparent at extreme positive velocities) as
can be seen in the bottom panel of Figure~\ref{fig:spec3}.

Another artifact appears as a low amplitude grid-like scanning
pattern in channel images with low-level emission.  This is due to
residual ripples in the bandpass of individual scans, which can cause
slight offsets between adjacent scans.  These are typically within the
noise, but because they are spatially correlated they can be seen in the 
rms image in Figure\ \ref{fig:rms}.

%----------------------------------------------------------
\section{Summary}
\label{sec:disc}
%----------------------------------------------------------

\begin{figure}
\centering
\includegraphics[width=3.3in]{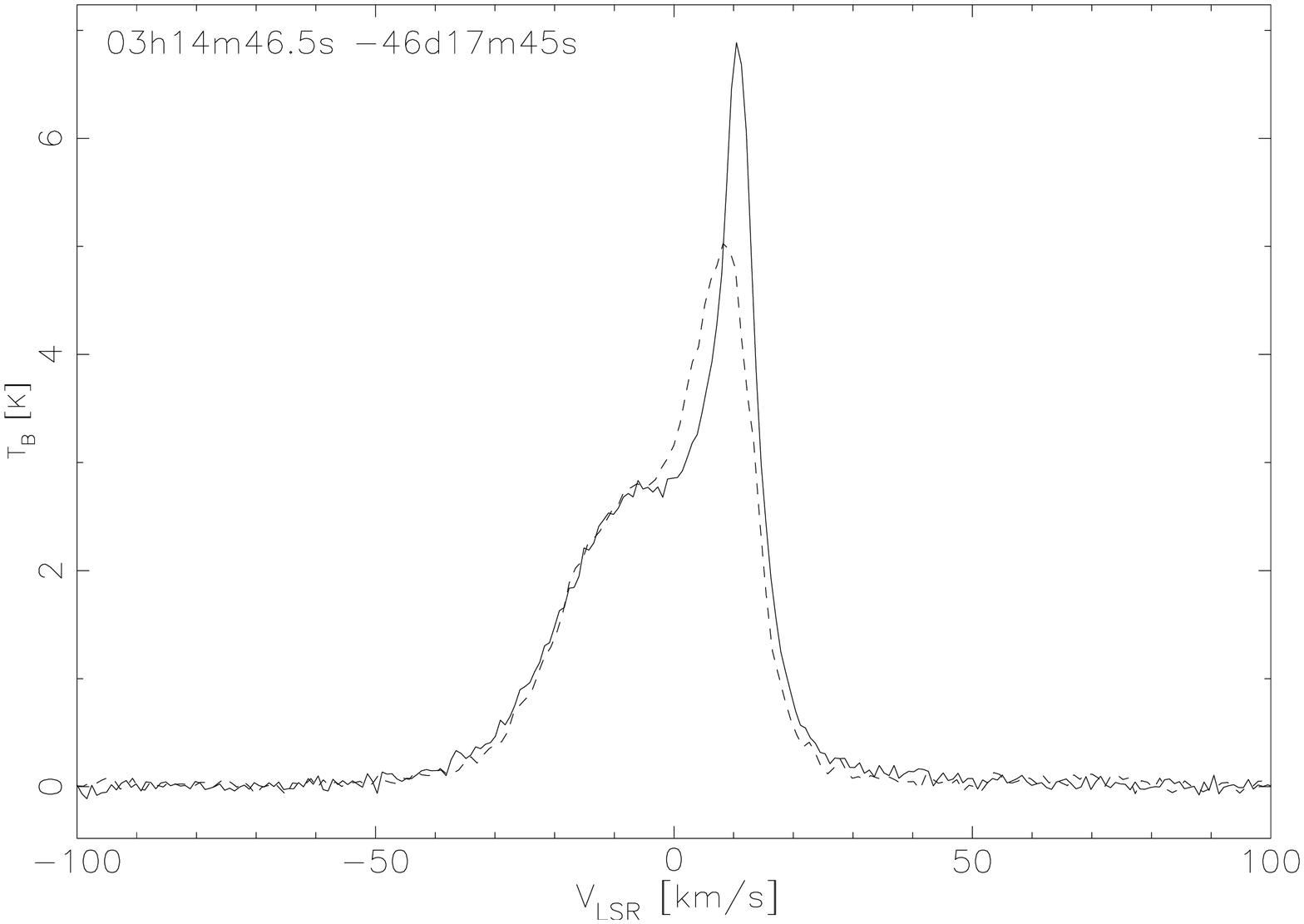}\\
\vspace{0.5cm}
\includegraphics[width=3.3in]{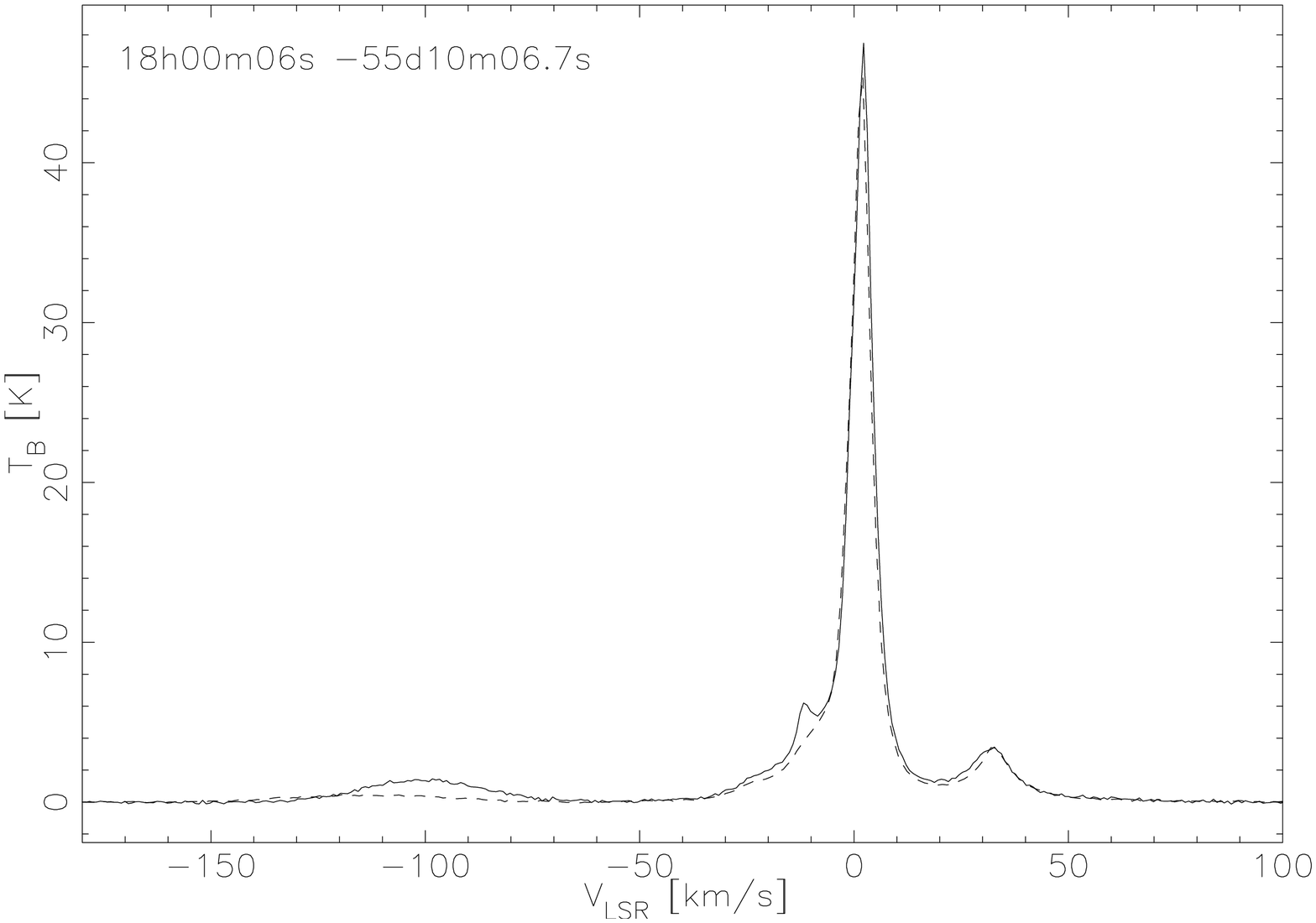}
\caption[]{Comparison of spectra from GASS (solid line) and LAB
  (dashed line) at the full GASS resolution towards two regions.  The general baseline
quality of GASS data is good and the higher angular resolution GASS spectra show 
features that are not apparent in LAB. \label{fig:spec_comparison}}
\end{figure}

The Parkes Galactic All-Sky Survey (GASS) is a high spectral and
angular resolution \HI\ line survey of the sky south of $\delta
=1\arcdeg$.  The first-stage data release includes a full cube at
$16\arcmin$ angular resolution, $1.0$ \kms\ spectral resolution and
$\sim 57$ mK rms noise over the velocity range $-400~{\rm km~s^{-1}}
\leq V_{LSR} \leq 500~{\rm km~s^{-1}}$.  Spectra from this cube and
several other data products are now publicly available at
\url{http://www.atnf.csiro.au/research/GASS}.  Table~\ref{tab:compare}
gives a comparison of GASS with other large-scale surveys of Galactic
\HI.  At southern declinations GASS is unsurpassed in sensitivity and,
outside of the Galactic plane, in angular and velocity resolution.
The improvement of GASS over the LAB survey is illustrated by the
spectra in Figure~\ref{fig:spec_comparison}, shown at the full angular
resolution of both surveys.

\begin{figure}
\centering
\includegraphics[angle=-90,width=3.2in]{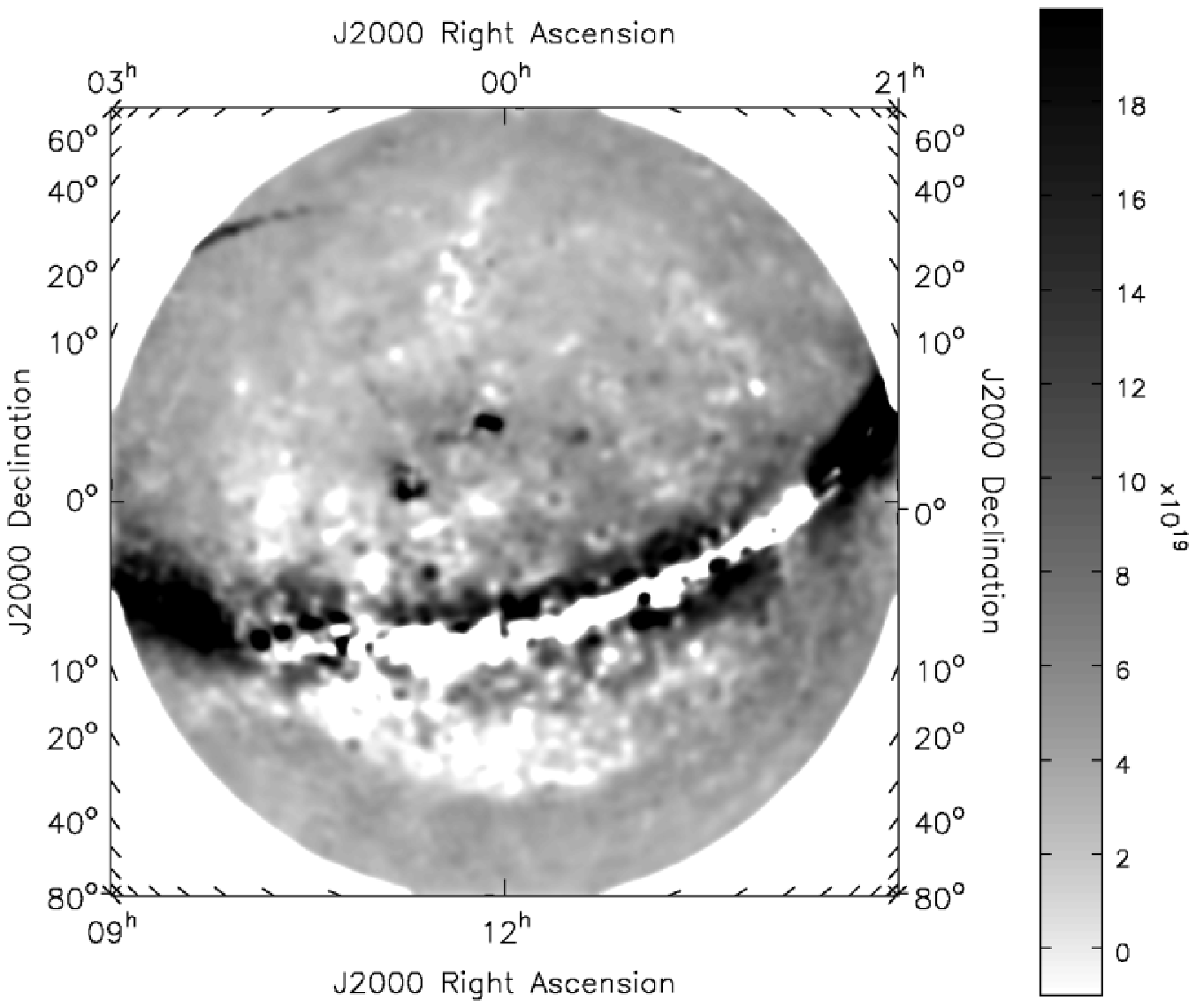}\\
\includegraphics[angle=-90,width=3.2in]{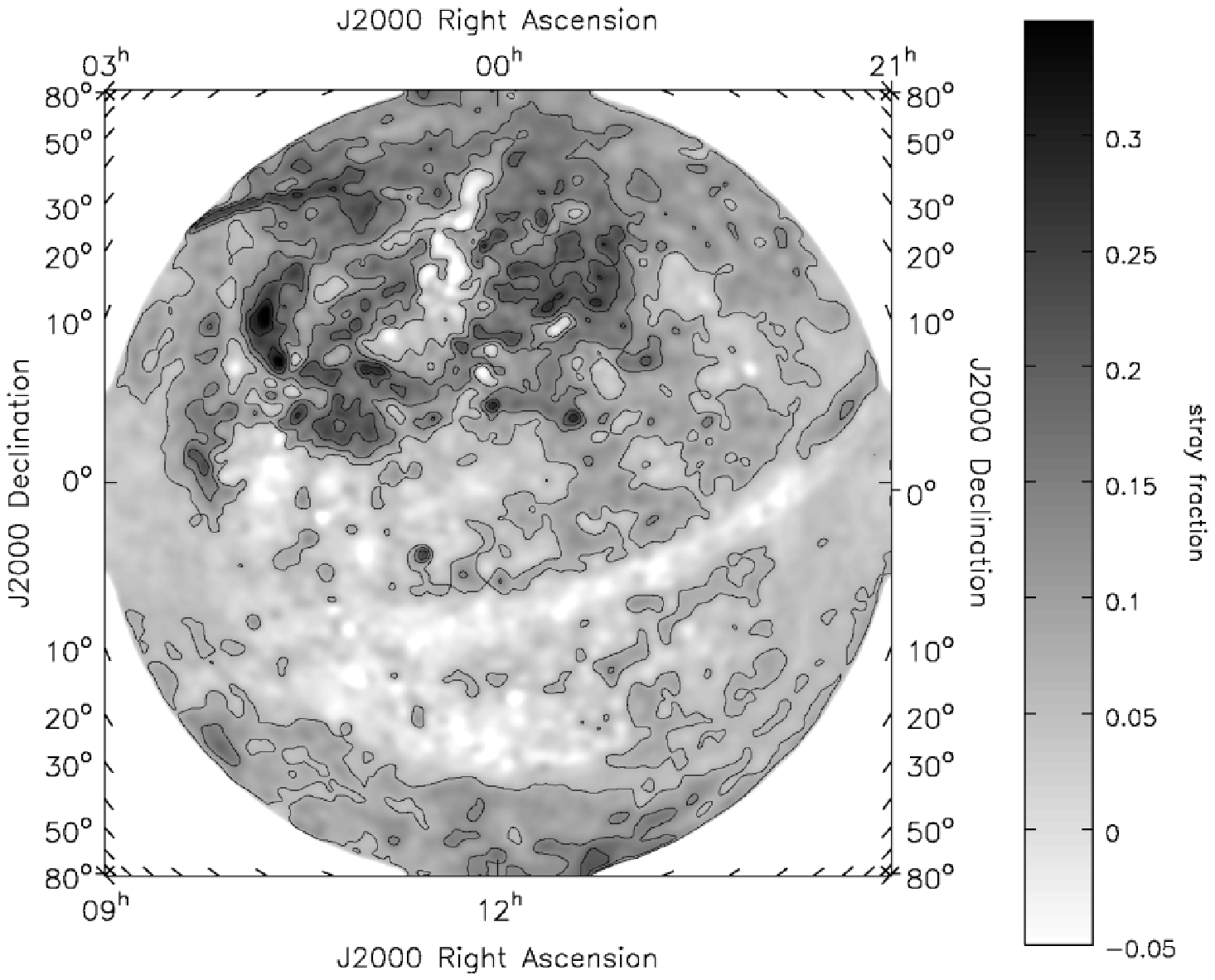}\\
\includegraphics[angle=-90,width=3.2in]{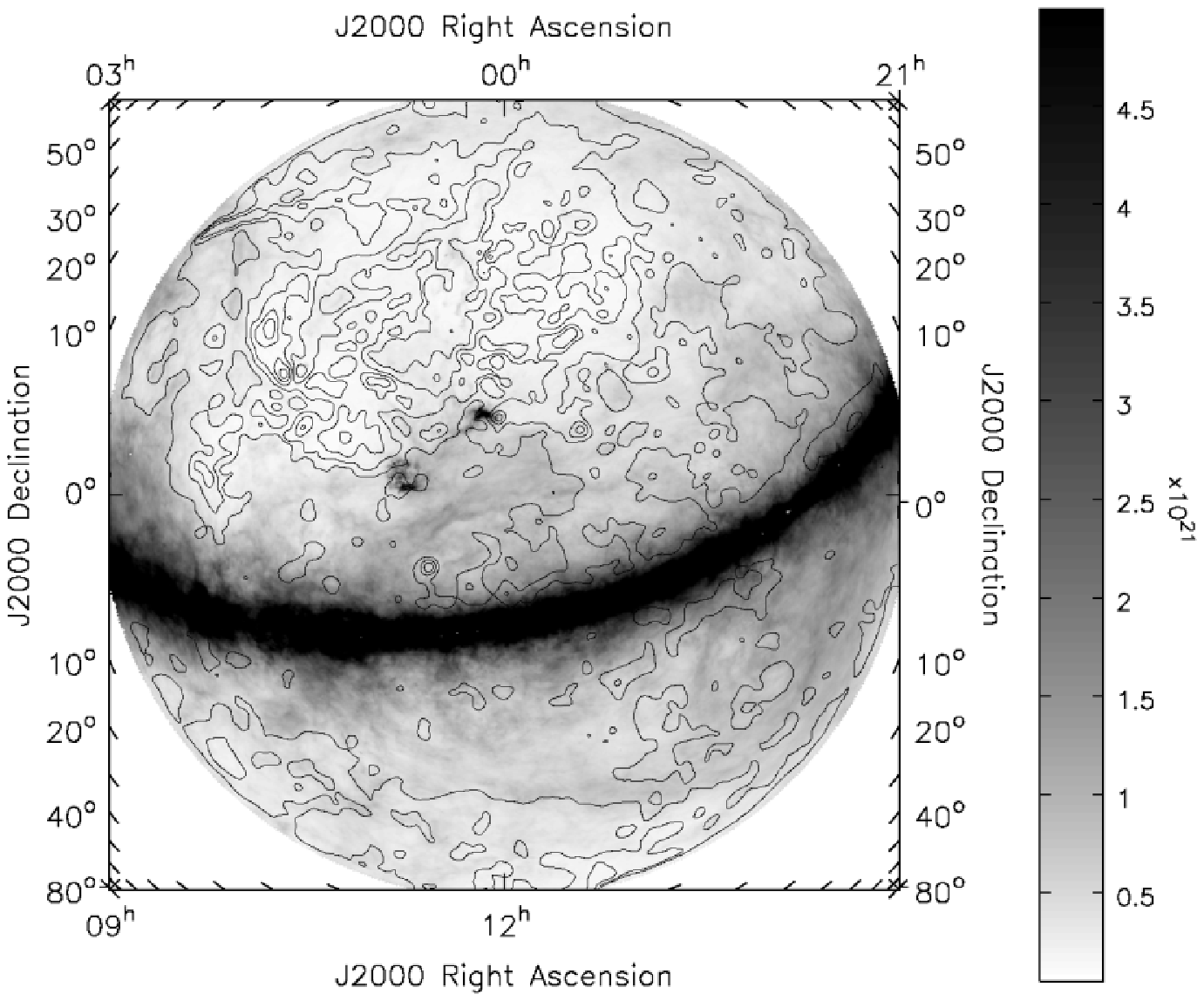}
\caption[]{({\em top}):  The amount of stray radiation in GASS expressed as
an equivalent $N_{HI}$.  The greyscale is linear from ($-1$ to
20)$\times 10^{19}~{\rm cm^{-2}}$ with a 
scaling power of $-0.5$.  The large negative and positive values towards the 
Galactic Plane are due to baseline problems and not stray radiation.
({\em middle}): Fraction of total column density estimated to be
  due to stray radiation. The greyscale is linear from $-0.05$ to
  $0.35$ as shown in the wedge.  The contours run from $-3$ to $5
  \times 0.065$.  ({\em bottom}): Total column density of GASS.  The
  greyscale goes from $5\times10^{19}$ to $5\times 10^{21}~{\rm
  cm^{-2}}$ with a scaling power of 0.5.  The contours are the same
  as in the middle panel.  The linear feature in the top left is due to an 
  artifact from regridding the LAB data.  All of the stray radiation data have
been smoothed to 2$\arcdeg\ $ resolution.
\label{fig:stray}}
\end{figure}

The current release of GASS is intended to provide data that are ideal
for study of high velocity ($|V_{LSR}|\gtrsim 100$ \kms) \HI, where
the effects of stray radiation are negligible, and small-scale
features, where the angular resolution is a significant advance.  A
complete catalog of HVCs and IVCs from GASS is in preparation by D.\ J.\
Pisano et al.\ (2009, in preparation).  Because the current data
release has not been corrected for stray radiation, care should be
taken when using it to derive global Milky Way properties, especially
at low column densities where there can be significant amounts of
stray radiation.  For users interested in total column densities
measured to extragalactic sightlines these measurements should be
considered in conjunction with the total fraction of the column due to
stray radiation (Figure~\ref{fig:stray}).

GASS data have already been used for several of the scientific areas
described in \S \ref{sec:intro}: Using data from the first year of
observations \citet{mcgriff06a} found that one of the largest Galactic
supershells, GSH 242--03+37, is in fact a chimney, with evidence for
breakout on both sides of the Galactic plane.  GSH 242--03+37 appears
capped by thin, clumpy filaments of \HI\ emission at heights of $z\sim
1.5$ kpc above the Galactic midplane.  Those authors suggested that
the clumpy filaments may be the precursor of halo clouds detected by
\citet{lockman02a}.

Ford et al.\ (2008) \nocite{ford08} used GASS data to extend our
knowledge of Galactic halo clouds by constructing a catalogue of
hundreds of clouds in a $720~{\rm deg^2}$ GASS pilot region centred on
$l=335\arcdeg$, $b=0\arcdeg$.  Though restricted to a small range of
Galactic longitudes, they found that the distribution of clouds is
significantly peaked at a Galactocentric radius of 3.75 kpc and that
the clouds are associated with loops and filaments consistent with a
chimney origin as suggested for GSH 242--03+37.  The forthcoming
complete catalog by H.\ A.\ Ford et al.\ (2009, in preparation) will
further explore the Galactic distribution of these clouds.

Finally, \citet{mcgriff08} have used GASS data to study the
interaction of an HVC in the Magellanic Leading Arm with the Galactic
disk.  They showed that the Leading Arm crosses the Milky Way disk at
a Galactocentric radius of 17 kpc, which is close to the interaction
region predicted in older models \citep[e.g.,][]{connors06,yoshizawa03}
of the Magellanic System but somewhat surprising given the revised
proper motions for the Large and Small Magellanic Clouds
\citep{vandermarel02,kallivayalil06b}.

As a new, sensitive survey of Galactic \HI, GASS is already producing
excellent results and will be a valuable database. Work is underway to
produce a stray-radiation corrected version of GASS for a second data
release.

\begin{figure}
\centering
\includegraphics[width=5.5in]{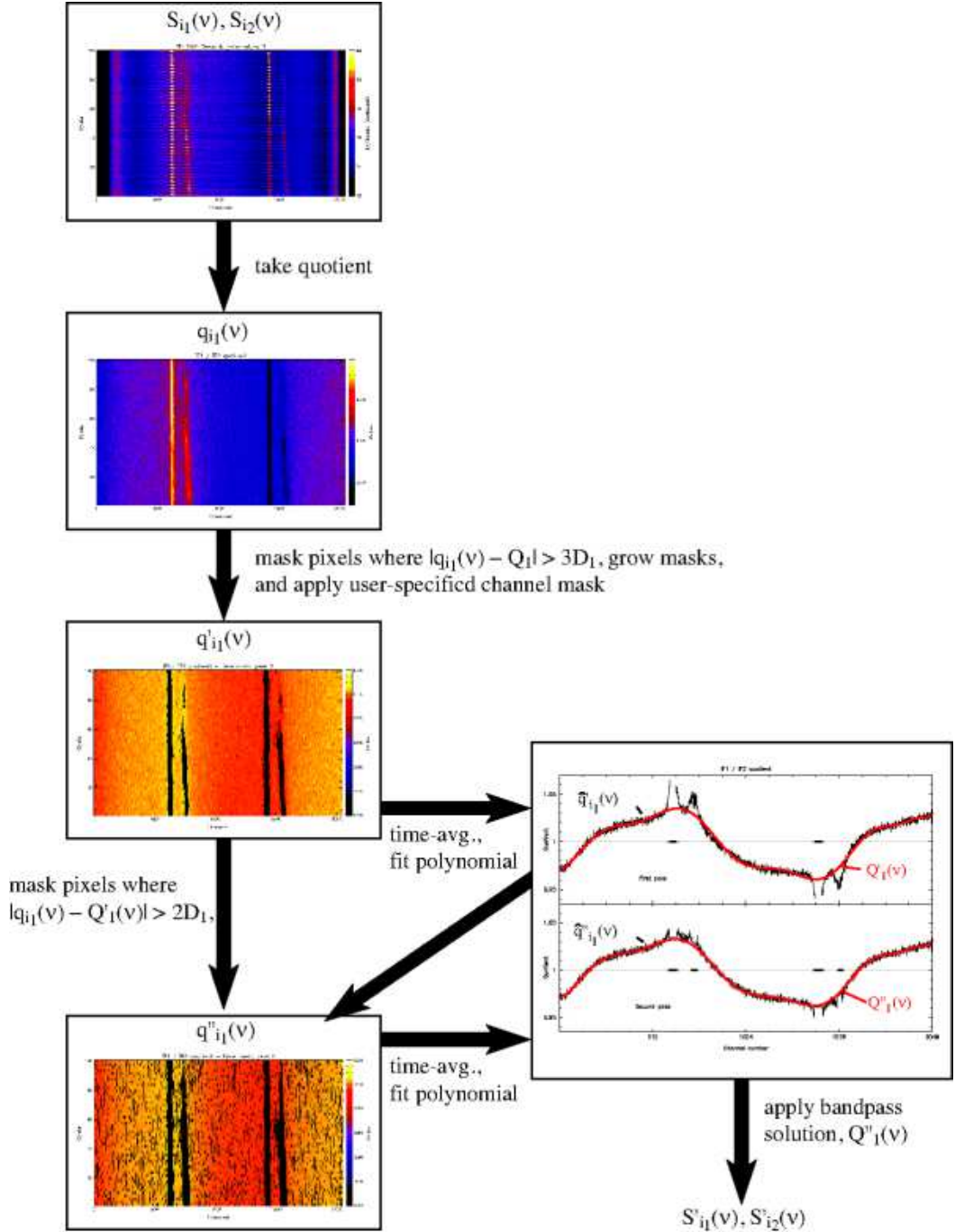}
\caption[]{Diagram outlining the bandpass solution steps outlined in
  \ref{subsubsec:bandpass}.  Each color panel represents data from a single
  scan, with time along the ordinate and channel number along the abscissa.
  The line plots show time-averaged spectra at two stages in the
  processing.  The black and grey solid lines at a quotient value of 1.0 represent the 
  channel masks; the former as derived from the time masks, the latter is 
  the user-defined channel mask.  Because the quotients for the two IFs
  are reciprocals of each other, we have only shown the quotients from the
  first IF.  The resulting solution can be applied to both IFs.   
  \label{fig:flowchart}}
\end{figure}

\acknowledgements We acknowledge the great dedication of the ATNF
staff at Parkes and Marsfield towards making the special observing
mode for this project available and supporting our observations.  We
especially thank Malte Marquarding for assisting with various software
challenges during the data analysis.  Thanks also to Bill Saxton for his
assistance in creating Figure~\ref{fig:flowchart}.  D.J.P.\ acknowledges 
partial support for this project from NSF grant AST0104439 and thanks 
the ATNF for its generosity and hospitality through the Distinguished 
Visitor program.  S.J.\ thanks the NSF for support through the Research
Experiences for Undergraduates program at NRAO.  T.M.\ acknowledges
the support of an ARC Australian Postdoctoral Fellowship
(DP0665973). P.K.\ and L.D.\ acknowledge support from Deutsche
Forschungsgemeinschaft, grant KA1265/5-1.  The Parkes Radio Telescope
is part of the Australia Telescope which is funded by the Commonwealth
of Australia for operation as a National Facility managed by CSIRO.

{\it Facilities:} \facility{Parkes ()}

%-------------------------------------------- 
%Bibliography
%--------------------------------------------
%\small 
%\clearpage
%\bibliographystyle{apj} 

%\bibliography{references.bib} %~mcg/tex/references.bib, bibtex file 
\normalsize

%---------------------------------------------
% Figures and Tables
%---------------------------------------------

\clearpage

\begin{deluxetable}{lc}
\tabletypesize{\scriptsize}
\tablecaption{Survey parameters
\label{tab:summary}}
\tablewidth{0pt}
\tablehead{\colhead{Parameter} & \colhead {Value}}
\startdata
Sky coverage & $\delta < 1\arcdeg$ \\
Integration time & $30~{\rm s}$ per spectrum\\
Central beam FWHM & $14.0\arcmin$ \\
Gridded Angular resolution & 16$\arcmin$ \\
Velocity range & $-400 < V_{LSR} < 500~{\rm km~s^{-1}}$ \\
Channel Spacing & $0.82$ \kms \\
Channel Width & $1.0$ \kms \\
$1\sigma$ $T_B$ noise & 57 mK\\
$3\sigma$ N$_{HI}$ sensitivity limit\tablenotemark{a} & $1.6\times
10^{18}~{\rm cm^{-2}}$\\ 
\enddata
\tablenotetext{a}{For $\Delta v  = 30$ \kms.}
\end{deluxetable}

\begin{deluxetable}{ccccccc}
\tabletypesize{\scriptsize}
\tablecaption{Survey Comparison \label{tab:compare}}
\tablewidth{0pc}
\tablehead{\colhead{Survey} & \colhead{Sky Coverage} & \colhead{Velocity Range}
& \colhead{Beamwidth} & \colhead{$\Delta$v} & \colhead{1$\sigma$ T$_B$} & 
\colhead{Reference} \\
&  & \colhead{(\kms)} & \colhead{($\arcmin$)} &  \colhead{(\kms)} 
& \colhead{(mK)} & }
\startdata
HIPASS-HVC & $\delta < 2\arcdeg$ & $-500 < V_{LSR} < 500$ & 15.5 & 26.4 & 8  & 1 \\
LAB & Entire sky & $-450 < V_{LSR} < 400$ & 36 & 1.3 & 70 & 2\\
SGPS & $253\arcdeg \le l \le 358\arcdeg$, & & & & & \\
 &  $5\arcdeg  \le l \le 20\arcdeg$, & $-100 < V_{LSR} < 126$\tablenotemark{a} & 2 & 0.8  & 1600 & 3 \\
 &  $|b|\le 1.5\arcdeg$ & & & & & \\
GALFA-HI & $-1\arcdeg < \delta < 38\arcdeg$ & $-700 < V_{LSR} < 700$ & 3.5 & 0.18  & $\leq100$ & 4 \\
GASS & $\delta < 1\arcdeg$ &  $-400 < V_{LSR} < 500$ & 16 & 0.82 & 57 & 5 \\
\enddata
\tablenotetext{a}{This is the velocity range covered by all SGPS cubes.  The 
actual range varies depending on the direction observed and has a maximal 
coverage of $-300 < V_{LSR} < 266$ \kms.}
\tablenotetext{1}{\citet{putman02}}
\tablenotetext{2}{\citet{kalberla05}}
\tablenotetext{3}{\citet{mcgriff05}}
\tablenotetext{4}{\citet{peek08}}
\tablenotetext{5}{This Paper}
\end{deluxetable}

\clearpage

\end{document}